\documentclass[pdftex,twocolumn,epjc3]{svjour3}          

\RequirePackage[T1]{fontenc}

\smartqed  

\RequirePackage{graphicx}
\RequirePackage{mathptmx}      
\RequirePackage{flushend}
\RequirePackage[numbers,sort&compress]{natbib}
\RequirePackage[colorlinks,citecolor=blue,urlcolor=blue,linkcolor=blue]{hyperref}

\usepackage{subfigure}

\usepackage{microtype}
\usepackage{multirow}

\journalname{Eur. Phys. J. C}

\begin{document}

\title{Perspectives of a single-anode cylindrical chamber operating in ionization mode 
and high gas pressure}

\author{R.~Bouet \thanksref{LP2I}  
	\and
	J.~Busto \thanksref{CPPM}
          \and
        V.~Cecchini \thanksref{e3,LP2I,SUBATECH}
        \and
        P.~Charpentier \thanksref{LP2I}  
         \and
        M.~Chapellier \thanksref{IJCLab}  
        \and
        A.~Dastgheibi-Fard \thanksref{LPSC}  
           \and
        F.~Druillole \thanksref{LP2I}  
           \and
        C.~Jollet \thanksref{LP2I}  
           \and
        P.~Hellmuth \thanksref{LP2I}  
             \and
        M.~Gros \thanksref{CEA}  
                 \and
        P.~Lautridou \thanksref{e1,SUBATECH}  
                  \and
        A.~Meregaglia \thanksref{e2,LP2I}  
               \and
        X.~F.~Navick \thanksref{CEA}  
               \and
        F.~Piquemal \thanksref{LP2I}  
       \and
        M.~Roche \thanksref{LP2I}  
       \and
        B.~Thomas \thanksref{e4,LP2I}  
    	}

\thankstext{e1}{e-mail: pascal.lautridou@subatech.in2p3.fr}
\thankstext{e2}{e-mail: anselmo.meregaglia@cern.ch}
\thankstext{e3}{Now at IFIC Valencia}
\thankstext{e4}{Now at CBMN Bordeaux}

\institute{LP2I Bordeaux, Universit\'e de Bordeaux, CNRS/IN2P3, F-33175 Gradignan, France\label{LP2I}
	\and	
	CPPM, Universit\'e d'Aix-Marseille, CNRS/IN2P3, F-13288 Marseille, France\label{CPPM}
         \and
         SUBATECH, IMT-Atlantique, Universit\'e de Nantes, CNRS-IN2P3, France\label{SUBATECH}
         \and
         IJCLab, CNRS/IN2P3, Paris, France\label{IJCLab}
           \and
         LPSC-LSM, CNRS/IN2P3, Universit\'e Grenoble-Alpes, Modane, France\label{LPSC}
            \and
         IRFU, CEA, Universit\'e Paris-Saclay, F-91191 Gif-sur-Yvette, France\label{CEA}
}

\date{Received: date / Accepted: date}

\maketitle

\begin{abstract}

As part of the R2D2 (Rare Decays with Radial Detector) R\&D, the use of a gas detector with a spherical or cylindrical cathode, equipped with a single anode and operating at high pressure, was studied for the search of rare phenomena such as neutrinoless double-beta decay. 
The presented measurements were obtained with a cylindrical detector, covering gas pressures ranging from 1 to 10~bar in argon and 1 to 6~bar in xenon, using both a point-like source of $^{210} $Po (5.3~MeV $\alpha$) and a diffuse source of $^{222}$Rn (5.5~MeV $\alpha$). 
Analysis and interpretation of the data were developed using the anodic current waveform.
Similar detection performances were achieved with both gases, and comparable energy resolutions were measured with both sources. 
As long as the purity of the gas was sufficient, no significant degradation of the measured energy was observed by increasing the pressure. 
At the highest operating pressure, an energy resolution better than 1.5\% full-width at half-maximum (FWHM) was obtained for both gaseous media, although optimal noise conditions were not reached.
\end{abstract}

\section{Introduction}
\label{Sec:intro}
The search for neutrinoless double beta decay ($\beta\beta0\nu$) is currently one of the main goals in particle physics. Several ton-scale projects have been internationally identified such as LEGEND~\cite{LEGEND:2022bzq}, CUORE~\cite{CUORE:2022uex}, or nEXO~\cite{nEXO:2017nam}. They aim to reach an effective Majorana neutrino mass, $m_{\beta\beta}$, of $\sim$10~meV, ruling out the inverted hierarchy region in the absence of any signals, employing at least one ton of material enriched in double beta decay isotopes. However, reaching the desired low-background level with a complex detector counting a large number of readout channels is not a trivial task. 
This weakness is also highlighted in other techniques, such as in time projection chambers (TPCs) utilising pressurized xenon gas to image the two expected beta traces~\cite{Next:2019}. Finally, the cost of the detector is also an important parameter to be considered. 

An alternative approach could be provided based on the concept of the Spherical Proportional Counter (SPC). 
This kind of radial TPC, first developed at CEA Saclay~\cite{Giomataris:2008ap}, was initially devoted to the physics of low-energy neutrinos, such as neutrino oscillations, neutrino coherent elastic scattering and supernova neutrinos detection~\cite{Giomataris:2003bp,Giomataris:2005fx}. In the last two decades, many important developments have been made.
The SPC's simple design, consisting of a large grounded sphere with a small central spherical anode as a unique readout channel,  minimizes the detector material budget. 
Furthermore, the SPC exhibits both a low energy threshold and a high gain, allowing the observation of the signal created by the production of a single primary electron.
This technology was chosen by the NEWS-G collaboration~\cite{Gerbier:2014jwa, NEWS-G:2017pxg, Savvidis:2016wei} for the search of light dark matter candidates in the mass range between 0.1 and 10~GeV.
Combined with a percent-level energy resolution, 
this detector could be an appealing device in the search for $\beta\beta0\nu$ decay. Monte-Carlo studies of a $^{136}$Xe-filled detector at 40~bar suggest a sensitivity covering the inverted mass hierarchy region~\cite{Meregaglia:2017nhx} with a ton scale detector. 
R2D2 R\&D began in 2018, with the hope of proving that an energy resolution of 1\% 
full width at half maximum (FWHM) was achievable at 2.458~MeV, corresponding to the transition energy 
($Q_{\beta\beta}$) of the $^{136}$Xe double beta decay. 
Preliminary studies reported the operation of the SPC in proportional mode with argon~\cite{Bouet:2020lbp, Bouet:2022kav} and xenon~\cite{Bouet:2023zyk}. 
As of 2021, a cylindrical proportional counter (CPC), using the same detector principle as the SPC, but with a wire as a central anode, was developed~\cite{Bouet:2023zyk}. The results showed a preference for a cylindrical cathode, therefore making the CPC the baseline design for the R2D2 detector. The main advances arise from three factors: the grounded anode, which results in noise almost independent of the high voltage applied to the cathode; the $1/r$ dependence of the electric field, inducing a stronger field away from the anode and minimizing the impact of electronegative impurities; and, finally, greater homogeneity of the field compared to that caused by the rod supporting the anode.
However, as discussed in Sec.~\ref{Sec:DA}, the main difference lies in choosing to operate the CPC in ionization mode rather than in proportional mode. This latter mode offers the significant advantage of reducing the bias demand (which was only available up to 6~kV), and eliminating the gain fluctuations inherent to the avalanche mechanism for energy resolution.

In this report we present the current state of this development. 
The experimental setup is discussed in detail in Sec.~\ref{Sec:exp}. Since gas purity is paramount to a correct detector operation, the gas recirculation system and the obtained purity are discussed in Sec.~\ref{Sec:gasrec} and ~\ref{Sec:gaspur}, respectively. The effect of temperature variation is addressed in Sec.~\ref{Sec:temp}. Signal modeling and processing are the subject of Sec.~\ref{Sec:sigp}, and data analysis is presented in Sec.~\ref{Sec:DA}.

\section{Experimental setup}
\label{Sec:exp}

\begin{figure}[t]
    \centering
    \includegraphics[width=\columnwidth]{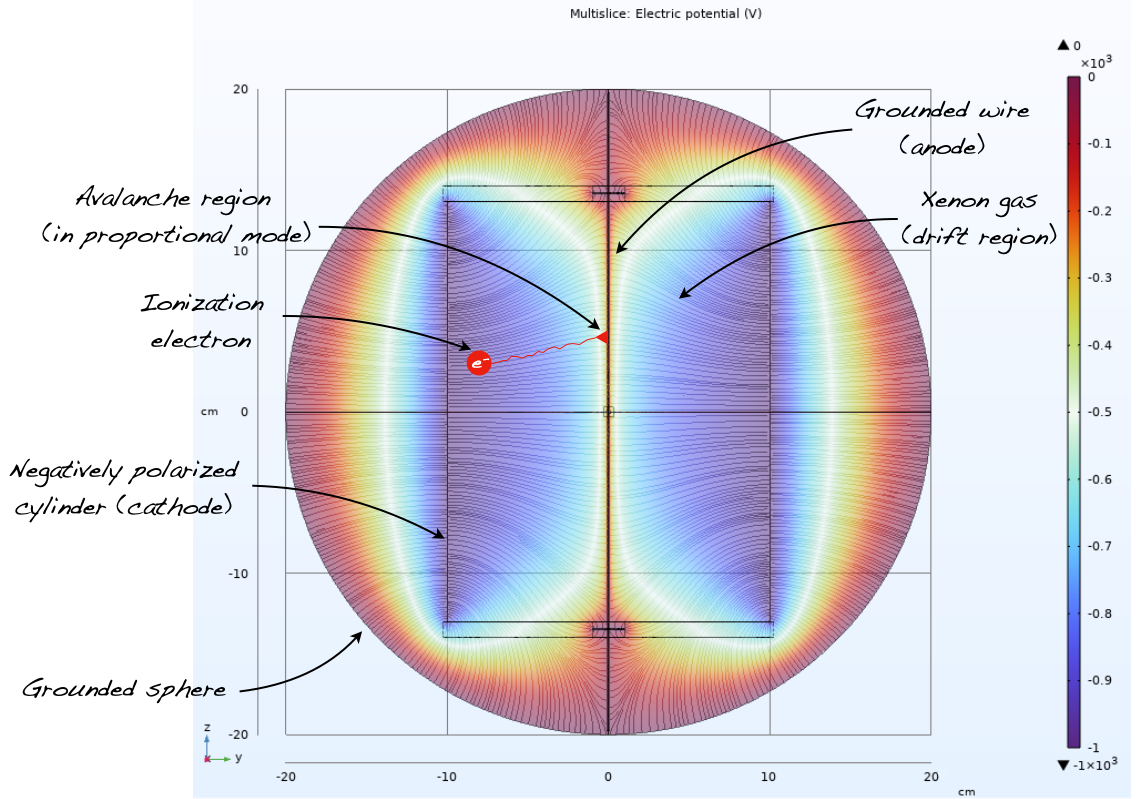}
    \caption{CPC working principle and electric field computed with COMSOL~\cite{COMSOL}.}
    \label{fig:Field}
\end{figure}

The detector is a CPC, consisting of three basic elements: a supporting structure, a cylindrical cathode, and a central wire. Moreover, in order to operate the CPC at pressures higher than atmospheric, a certified vessel withstanding the pressure is required.
The CPC working principle is identical to any gas proportional counter: ionisation tracks traversing the gaseous medium deposit energy and produce ionisation electrons. An electric field, produced by a negative high-voltage applied on the cylinder (cathode), drifts the electrons towards the grounded central wire (anode), which is used to read out the induced signal. Depending on the value of the electric field, the electrons could undergo an avalanche process near the wire (proportional mode) or simply be collected with no amplification (ionisation mode). The electric field, computed with COMSOL software~\cite{COMSOL}, as well as the detector working principle can be seen graphically in Fig.~\ref{fig:Field}.

\subsection{High pressure vessel}

\begin{figure}[t]
    \centering
    \includegraphics[width=0.7\columnwidth]{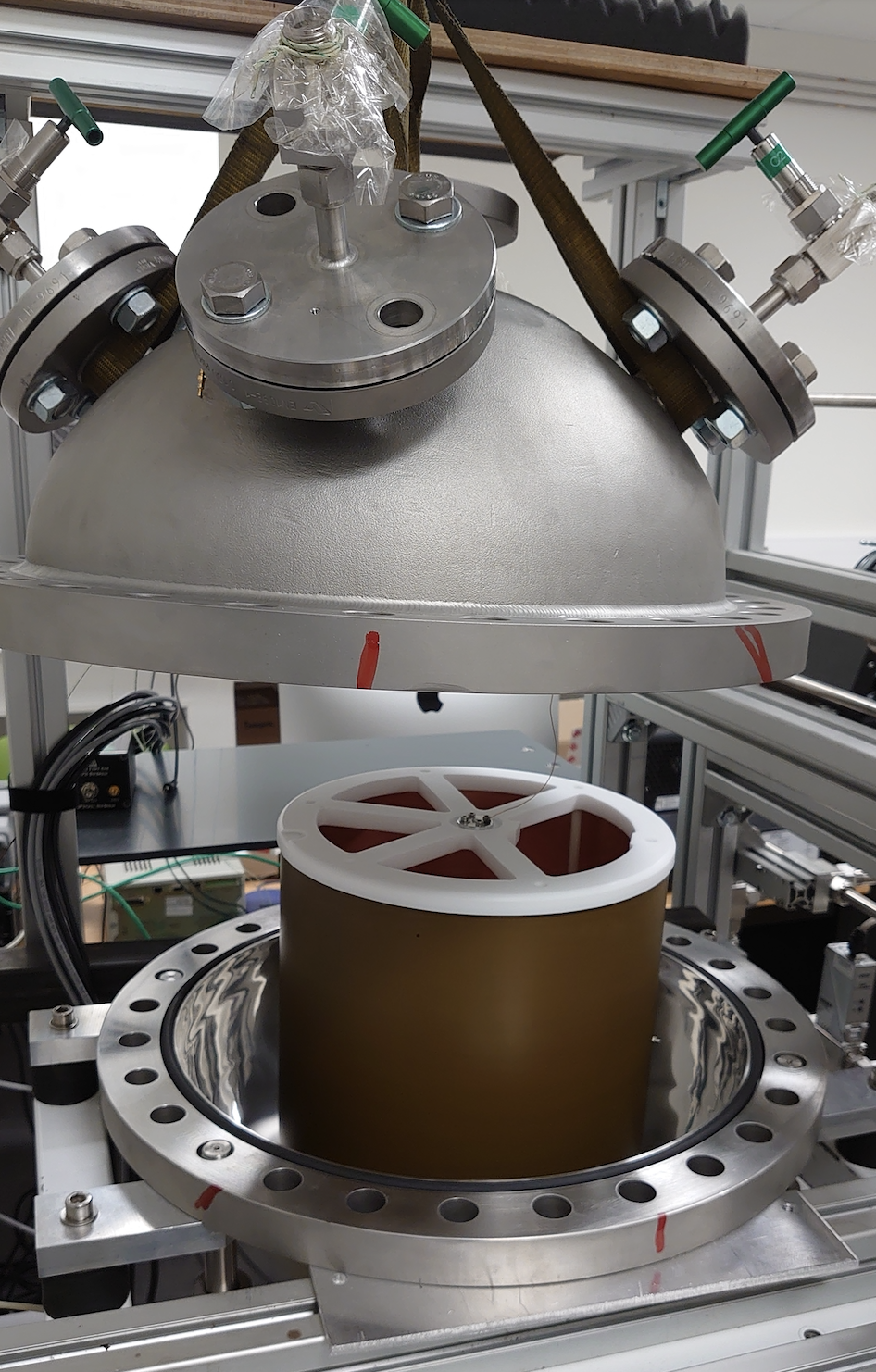}
    \caption{CPC set in the spherical vessel before final assembly. The wheel structure (white Teflon) supports the cathode (epoxy-copper type). The anode is attached to the central insert.}
    \label{fig:Detector}
\end{figure}

 To avoid additional costs, the available spherical vessel (of 20~cm radius and volume of approximately 33~l), built in the first R2D2 phase and tested as SPC, was used as the high pressure vessel to host the CPC. The cathode dimensions (with height and radius of 27~cm and 11~cm, respectively) were therefore chosen in order to fit inside the existing vessel. To hold the counter vertically inside the sphere, two specifically designed pieces were attached to both end-caps of the supporting structure. A picture of the system (with an epoxy-copper cathode), set in the bottom hemisphere of the spherical vessel, is shown in Fig.~\ref{fig:Detector}. 

\subsection{CPC supporting structure}
 
\begin{figure}[t]
    \centering
    \includegraphics[width=\columnwidth]{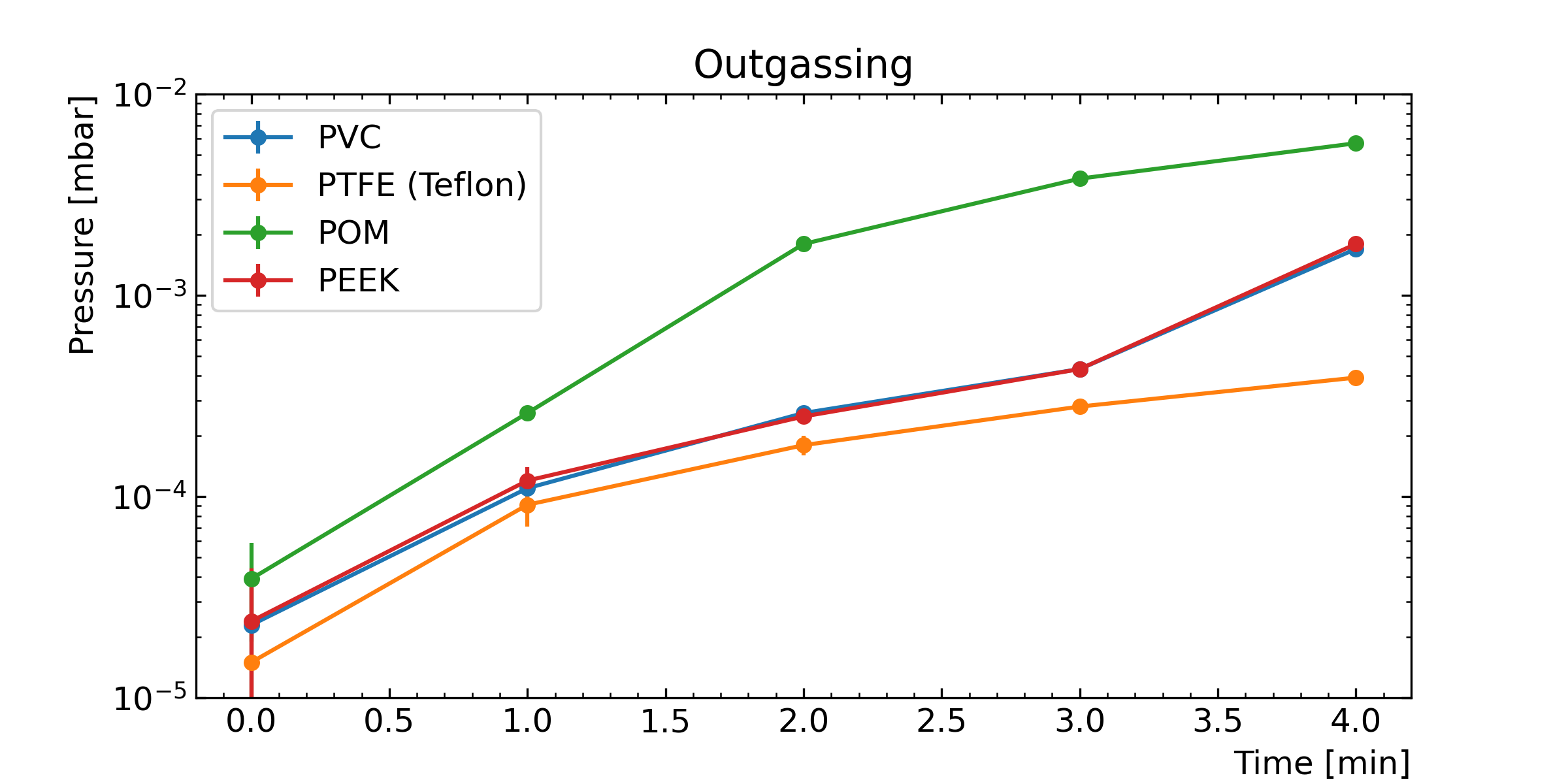}
    \caption{Evaluation of outgassing for some plastics polymers, potentially used for the CPC structure.}
    \label{fig:OutgassingMaterials}
\end{figure}

The skeleton consists of three cylindrical pillars holding two end-caps together. These end-cap pieces have the shape of a spoked wheel (to provide stability to the detector) surrounded by a groove (to hold the cathode in place).
In order to ensure an optimal gas purity, as discussed in the following sections, all the materials were selected to minimize the outgassing. Several plastics polymers were considered, such as PVC (polyvinyl chloride), Teflon (PTFE polytetrafluoroethylene), PEEK (polyether ether ketone) and POM (polyoxymethylene), and their outgassing was measured. The procedure involved inserting the material in a dedicated vacuum chamber and pumping for 30 minutes. Afterward, the pump was turned off, and the subsequent pressure increase in the vacuum chamber was monitored over a few minutes to assess outgassing. Teflon was chosen as the optimal material for the CPC structure, as indicated in the results presented in Fig.~\ref{fig:OutgassingMaterials}.


\subsection{CPC cathode}

\begin{figure}[t]
    \centering
    \includegraphics[width=\columnwidth]{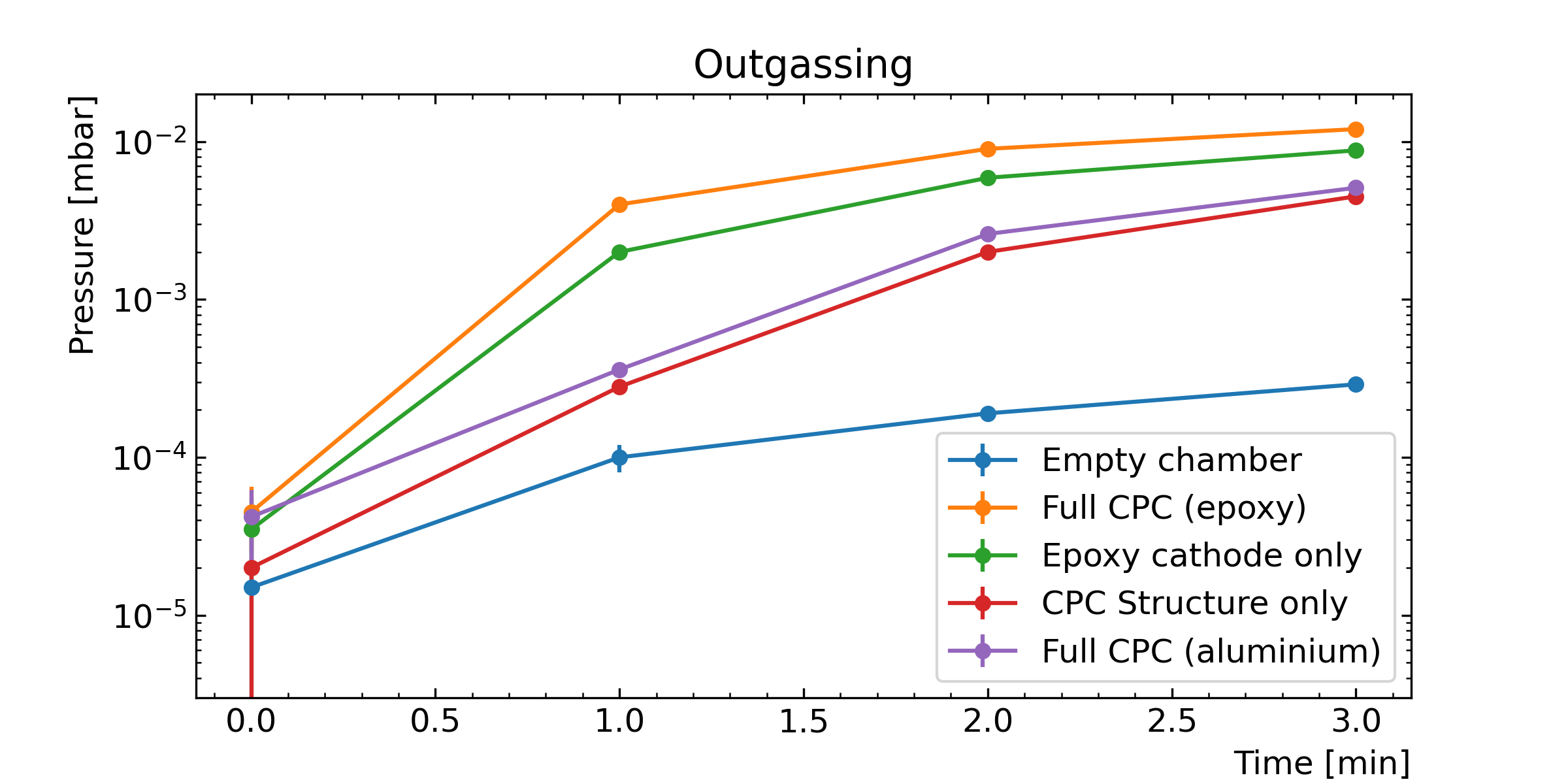}
    \caption{Evaluation of outgassing for different parts of the CPC and two cathodes: aluminium and copper covered with epoxy.The ``Empty chamber'' curve (blue line) labels the measurement of the vacuum chamber only, with no material inside.}
    \label{fig:Outgassing}
\end{figure}

The cathode choice was also motivated by the lowest possible outgassing.
As part of the CPC skelton, it must to be thin enough to be bent and inserted in the end-cap grooves, whilst assuring rigidity to the detector. A sheet of aluminium 200~$\mu$m thick, immediately available, was used as a cathode. The distance between the CPC wall and the grounded spherical vessel is large enough to safely apply high-voltage at the level of 5~kV without sparking. An insulating layer on the outer side of the cathode would be desirable, but this greatly increases the risk of outgassing. Two copper cathodes covered with kapton and epoxy as insulating materials were tested, revealing important outgassing: this is likely due to the material itself (epoxy) or to the process used to attach the two materials (glue between kapton and copper). The outgassing results of different components of the CPC and different cathodes are shown in Fig.~\ref{fig:Outgassing}: the full CPC with an aluminium cathode is almost identical to that obtained with the CPC structure only, proving the aluminium cathode is not a source of oxygen or other electronegative impurities. This cathode was therefore chosen for the presented measurements. The final setup was baked at 100 degrees Celsius for one week in a vacuum chamber to reduce the outgassing of the materials during operation.

\subsection{CPC central wire}

The diameter of the central wire is a principal parameter and is selected according to the chosen detector operation mode ({\it i.e.} ionization or proportional). In the case of ionization, a thicker wire ensures a larger electric field far from the wire, and the possibility to increase the electron's drift velocity, without starting an avalanche near the anode. In proportional mode, a thinner wire allows reaching the multiplication region with a relatively small high voltage ({\it i.e.} below 10~kV).
The measurements presented were conducted with two different anodes: a tungsten wire with a 50~$\mu$m diameter, and a metallic rod with a 1.2~mm diameter.
The wire/rod is connected to a feed-through on the top of the vessel and read through the electronics chain and DAQ to record the signal. The high voltage is applied to the CPC wall via a wire connected to a feed-through at the bottom of the vessel.
The coupling between the signal and high-voltage is minimized (at the level of the irreducible anode-cathode capacitance) ensuring the weakest dependence of the high-voltage on the noise. This choice was a major improvement over previous prototypes~\cite{Bouet:2020lbp,Bouet:2023zyk}, where the increase in high-voltage at high pressure subsequently increased the electronic noise, limiting the energy resolution.

\subsection{Read-out electronics}

The electronics chain consists of a low-noise charge preamplifier sending its waveform to a CALI card (ADC) which was controlled by the SAMBA acquisition software~\cite{EDELWEISS:2017lvq}. 
The integrator output signal is continuously sampled at 2~MHz. An internal digital amplitude threshold discriminator commands the capture of the waveform of interest for a duration of 3 ms.
The first 1.5 ms is devoted to estimating the baseline of the event, while the second half of the time window, 
containing the triggering physical transient, undergoes further digital processing (see Sec.~\ref{Sec:sigp}). 
The event trigger was carried out by positioning a fixed digital threshold on the amplitude of the wave train.
For this program of operation in ionization mode, the charge-sensitive low noise and high gain amplifier ORTEC 142PC was adopted to read the anode sensor. 

\subsection{Calibration source}

A $^{210}$Po source, emitting $\alpha$'s at 5.3~MeV, is used  to assess the detector energy resolution. The polonium is deposited on a silver plate of $0.6 \times 0.6$~cm$^2$, and positioned on the outside of the cathode behind a hole of 1~mm radius, through which the $\alpha$ particles enter the CPC active volume. The activity of the source was roughly 10~Bq, corresponding to a rate of 0.8 events per second in the detector.

%


\section{Gas recirculation system}
\label{Sec:gasrec}

As with all detectors utilising significant drift distances of the primaries ionisation electrons, a clean gas is crucial to limit the effects of electronegative impurities, such as oxygen or water.
These molecules can capture the ionization electrons, which then migrate towards the anode, thus reducing the charge collected.
In the case of noble liquid detectors, several solutions exist to guarantee a purity at the level of tens of ppt (part per trillion) or, as more commonly measured, an ``electron lifetime'' of several milliseconds.
Currently, the most effective purification method, used by the XENON collaboration, is based on cryogenic filters built from copper~\cite{Plante:2022khm}: oxygen impurities react with copper (2Cu+O$_2$ $\to$ 2CuO) and become permanently bound to its surface. Other purification processes could be applied such as the spark discharge technique~\cite{Akimov:2017gxm}, or cold and hot getters~\cite{BOLOTNIKOV1996619,Dobi:2010ai,Vignoli:2015jxa}. The use of commercial getters does not require cryogenic and distillation infrastructures, and are much cheaper compared to the cryogenic filter. However, the best achievable gas purity is then reduced to the ppb (part per billion) level.

An additional aspect which is normally taken into account in gas/liquid purification is the radon contamination~\cite{XENON:2020fbs}. Radon is the dominant source of background for experiments searching for rare events and it is typically removed through distillation columns~\cite{XENON100:2017gsw}. However, in the present phase of the R2D2 R\&D, there is no low background requirement and the presence of radon does not represent an issue. On the contrary, as will be discussed in the following, radon will be exploited as a diffuse source of calibration.   

\begin{figure}[t]
    \centering
    \includegraphics[width=0.9\columnwidth]{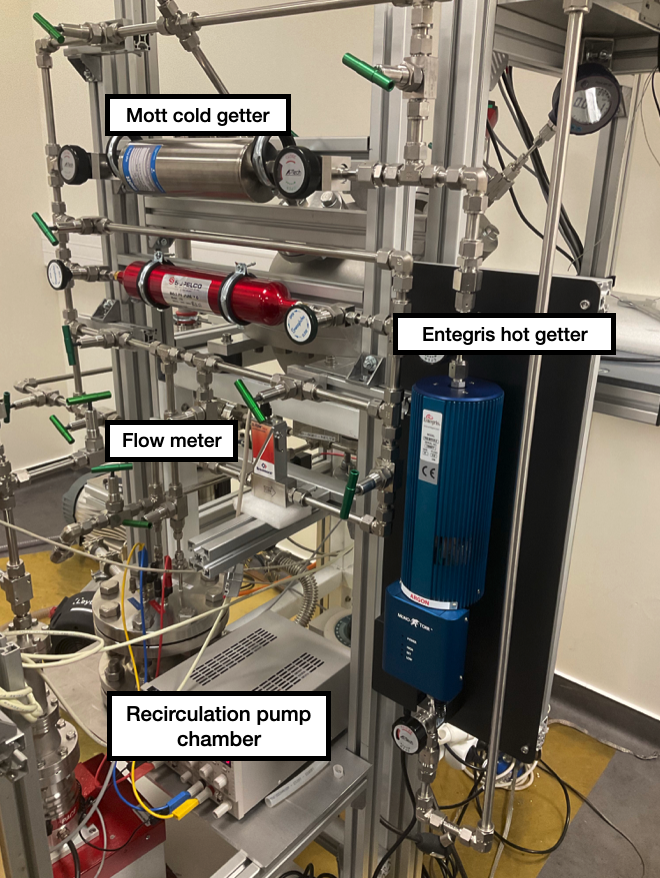}
    \caption{Recirculation system including cold and hot getter, as well as the flow meter and the recirculation pump chamber.}
    \label{fig:PicsRecirculation}
\end{figure}

In order to operate the R2D2 prototype with sufficient gas purity conditions, a combination of cold and hot getter was chosen. A PS3-MT3-R-2 MonoTorr Entergris gas purifier was used as hot getter, granting outlet impurity levels for oxygen reduced to ppb levels. The hot getter was coupled to a Mott micro-bulk gas purifier (MGP-30-125-IG-106-1.5NM-V3-V) cold getter, located upstream, to provide a first purification stage and increase the lifetime of the hot getter filtering components. This decision was mainly motivated by the allocated R\&D budget, which was not sufficient to have purification based on cryogenic filters.
Nonetheless, in terms of electronegative impurities, the contamination at the level of few~ppb is sufficient to operate the detector up to 10~bar. The 10~bar limit is determined by the maximal operating pressure of the getter.

The gas recirculation system is the second component ensuring the detector operates under the required conditions.
The gas was recirculated using a KNF pump (NPK03KVDC-B4) installed in a dedicated chamber to prevent leaks and withstand the gas pressure. The gas flow, measured by a Bronkhorst El-FLOW flowmeter, can be adjusted by varying the voltage applied to the pump. The two limiting factors are a maximal allowed flow of 10~l/m in the hot getter, and a limit on the rotation speed of the pump which induces vibrations on the setup increasing the noise. The latter limit depends on the gas pressure and was set, in practice, by requiring the same noise level when
taking data with and without the recirculation system. A picture of the recirculation setup can be seen in Fig.~\ref{fig:PicsRecirculation}.

\begin{figure}[tp]
    \centering
    \includegraphics[width=\columnwidth]{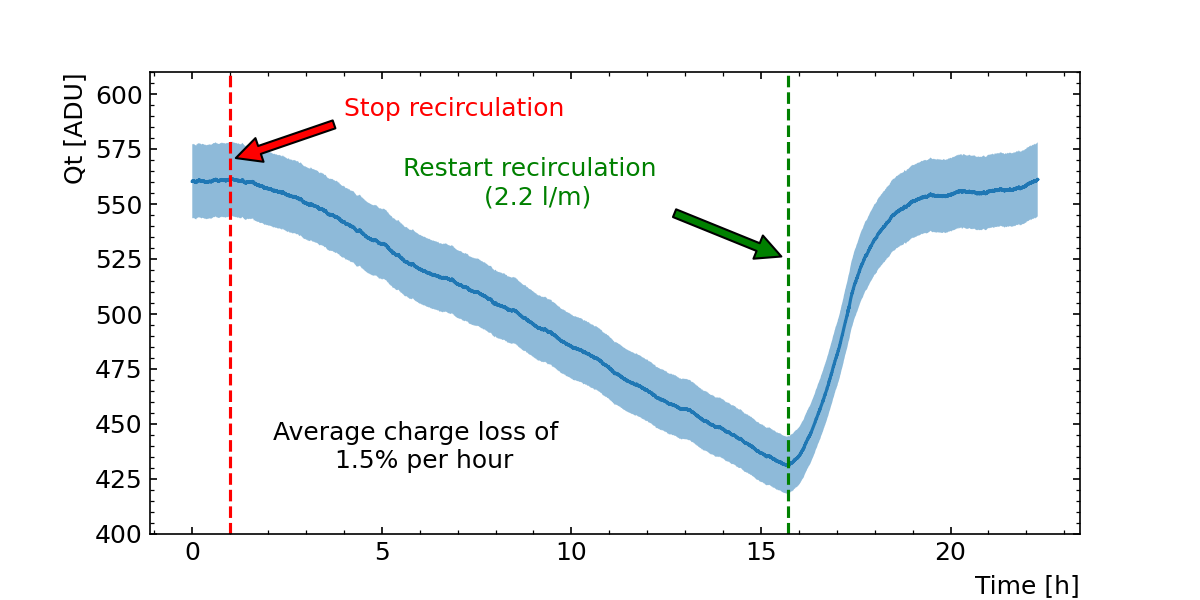}
    \caption{Time evolution of the reconstructed average charge, corresponding to 5.3~MeV $\alpha$'s, for data taken at 3~bar with Xe. The stop and restart of the recirculation system are marked by the red and green arrow, respectively.}
    \label{fig:Recirculation}
\end{figure}

To establish the need of a persistent recirculation while taking data, the gas purity degradation was assessed as a function of time. For this, a dedicated run was taken with the detector filled with xenon at 3~bar, exploiting the $^{210}$Po calibration source on the cathode. The mean reconstructed charge corresponding to $\alpha$ events was monitored over the whole run. At the beginning of the run the detector was in a stable condition, with the gas recirculating at the speed of 2.2~l/m for a few days. 1~h after the run began, the recirculation was stopped, and a constant reduction in the collected charge of about 1.5\% per hour was observed. Such a reduction demonstrates that a permanent recirculation (or at least regular) is needed during data taking to guarantee an optimal gas quality. 16~h after the run start, the recirculation was restarted, and the same reconstructed charge achieved at the beginning of the run was retrieved after only $\sim4$~h.  These observations are summarized in Fig.~\ref{fig:Recirculation}.


\section{Gas purity}
\label{Sec:gaspur}

\begin{figure}[t]
    \centering
    \includegraphics[width=\columnwidth]{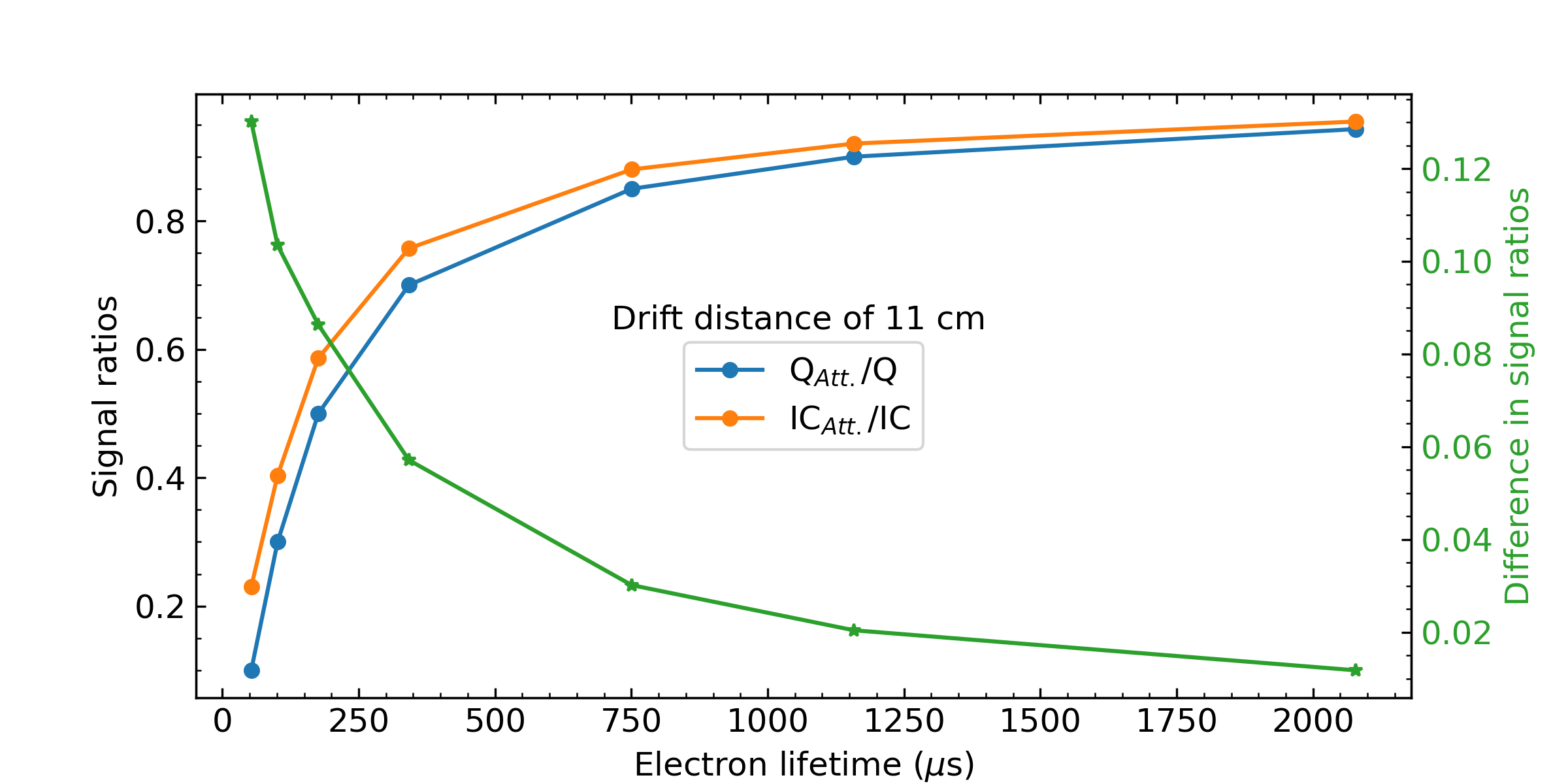}
    \caption{Ratio between attenuated and total signal for the induced current (orange line) and for the total collected charge (blue line). The green line shows the difference between the two signal ratios.}
    \label{fig:Attenuation}
\end{figure}

Gas purity effect is usually summarized in terms of ``electron lifetime''. This is a typical estimator of the contamination in terms of electronegative impurities in noble liquid/gas experiments. Indeed, in such TPCs, the number of charges present in the drift space
(and finally the collected charge) for a given energy deposit, depends on the electron drift time (and therefore on the position of the primary ionization along the drift direction).
For a deposited charge producing a total number of $N(0)$ electrons at time $t=0$, the collected number of electrons $N(t)$ after a drift time $t$ should follow the relation~\cite{Plante:2022khm}: 
 \begin{equation}
 \label{eq:1}
 N(t)= N(0) e^{-\frac{t}{\tau}} ,
\end{equation}
where $\tau$ is the so-called ``electron lifetime''. A long electron lifetime implies a smaller recombination during the drift, and thus a medium with a small contamination of electronegative impurities.
It must be noted that the impact of the electron lifetime on the reconstructed signal is different for detectors operated in proportional and ionisation mode and, from the point of view of the gas purity, operating the detector in the ionization regime should provide a slight advantage. The ionization signal begins as soon as the primary electrons 
are in motion, whereas the proportional signal, formed by the backwards drift of the produced ions, depends instead on the number of primaries which enter in the avalanche region, {\it i.e.} those that reach the anode. The effect of a loss of charges during 
migration is presented for the two regimes in Fig.~\ref{fig:Attenuation}. The calculation is carried out in the extreme case of a point-like interaction at the cathode position. 
The ionization signal ratio with and without charge losses ({\it IC$_{Att}$/IC}) is compared to the ratio of charges ({\it Q$_{Att}$/Q}) for different values of the electron lifetime.
The curves show that ionization mode presents better immunity to this effect and that the constraints on the gas purity could be less stringent. Furthermore, in ionization, this effect could likely be tuned by adjusting the induced signal shape, thereby accounting for the loss of primaries during transport (see Sec.~\ref{Sec:sigp}).

\begin{figure*}[tp]
    \centering     
   	 \subfigure[\label{fig:PurityPressure}]{\includegraphics[width=\columnwidth]{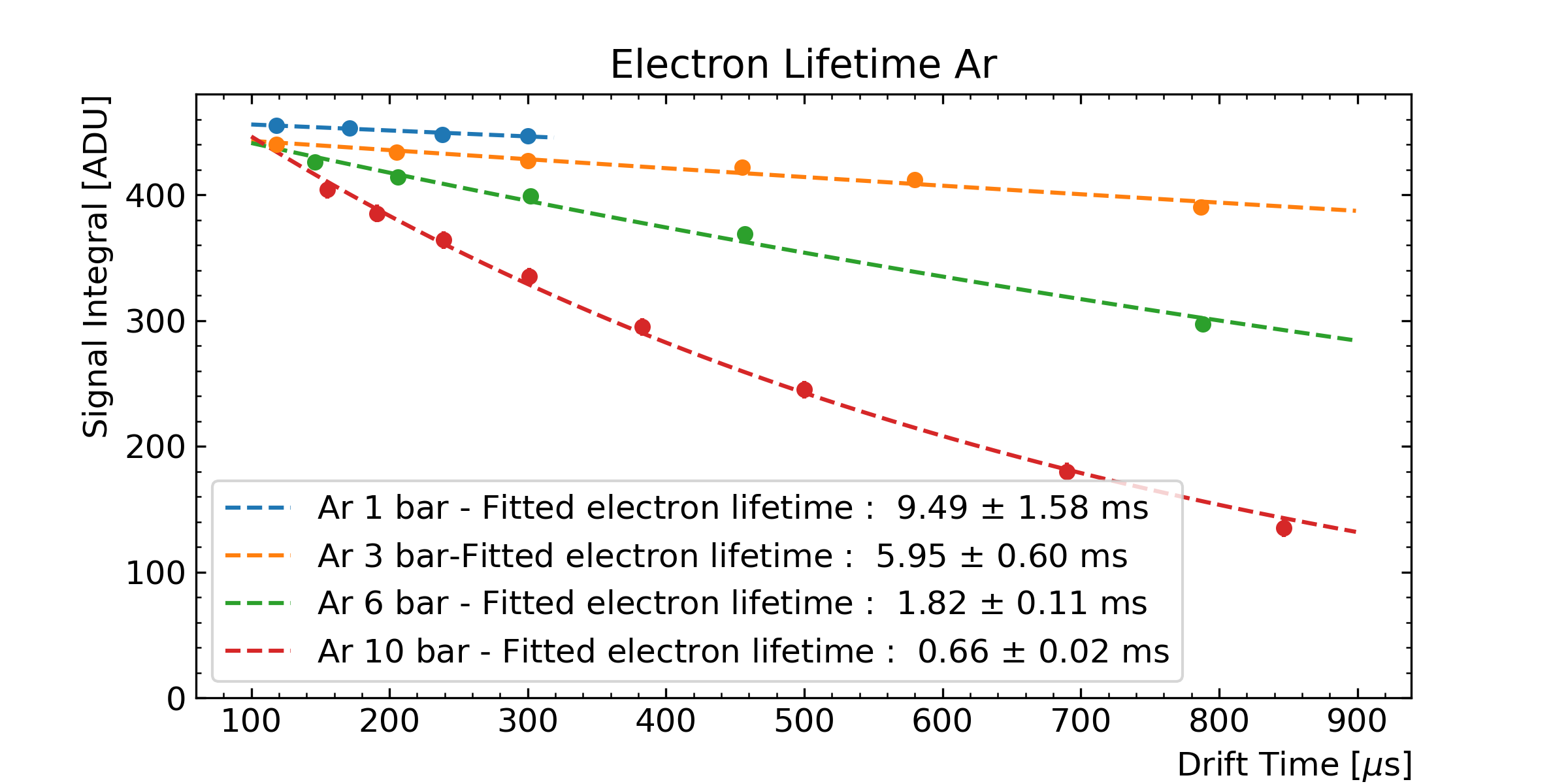}}
    	 \subfigure[\label{fig:PurityPressureXe}]{\includegraphics[width=\columnwidth]{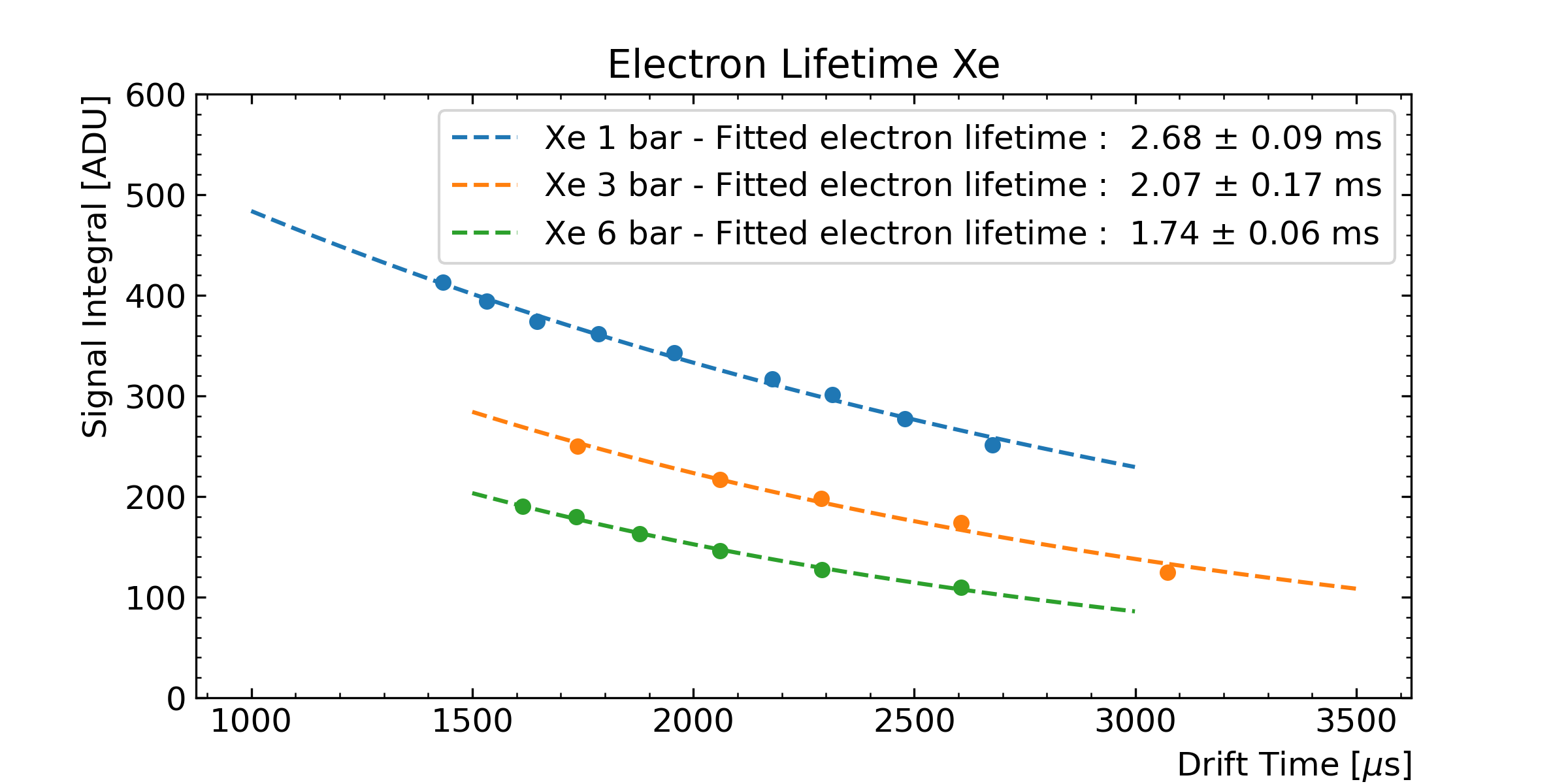}}
    \caption{Reconstructed charge as a function of the electron drift time (obtained from 2023 GARFIELD++ release) in argon (\ref{fig:PurityPressure}) and xenon (\ref{fig:PurityPressureXe}). The different colours represents different gas pressures: 1~bar (blue), 3~bar (orange), 6~bar (green) and 10~bar (red). The dashed line represents the fit done using Eq.~\ref{eq:1} and the obtained electron lifetime is shown on the plot.}
    \label{fig:PurityPressureAll}
\end{figure*}

\begin{figure}[b]
    \centering
    \includegraphics[width=\columnwidth]{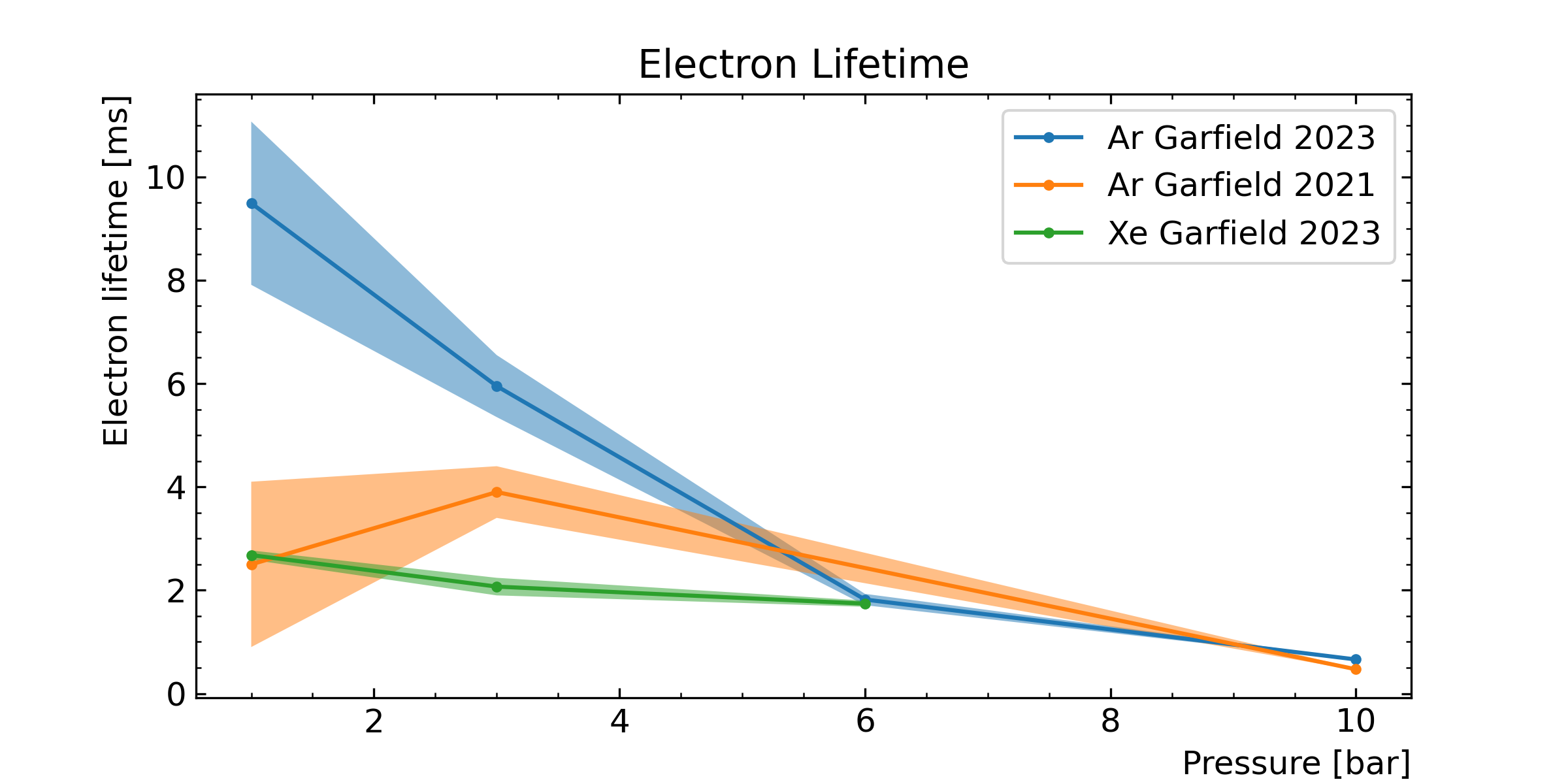}
    \caption{Electron lifetime at different pressures for argon, based on the drift time computed with 2023 GARFIELD++ release (blue) and  2021 release (orange), and xenon  based on the drift time computed with 2023 GARFIELD++ release (green). }
    \label{fig:PurityVsPressure}
\end{figure}

 In R2D2, the $^{210}$Po source is located on the cathode and it is not possible to vary its position radially along the drift direction. Another way to vary the drift time is to change the high-voltage applied on the cathode, and therefore the electric field, leaving the $\alpha$ source position unchanged. The two methods are not exactly equivalent, since the modification of the electric field to change the electron drift time could also modify the interactions between drifting electrons and electronegative impurities, and therefore the attachment. Considering such an effect, and that R2D2 is operated in ionisation mode, the exponential fit of Eq.~\ref{eq:1} is only an approximation for the electron lifetime evaluation. 
 Considering the method's limitations, measurements of the signal integral were taken at various high-voltages, \textit{i.e.} different electron drift times.
However, in the present experimental setup, there is no independent measurement of the drift time since the $\alpha$ emission time is unknown, and no trigger based on the scintillation light is present.
As is well established for chambers working in ionization regime, an experimental measurement of the drift time (and therefore the position of the interaction) can instead be deduced from the estimate of the signal duration (see Sec.~\ref{Sec:sigp}). 
For validation, the GARFIELD++~\cite{Veenhof:1993hz,Veenhof:1998tt} simulation toolkit was used to evaluate the drift time of electrons from the cathode to the central anode at different pressures and high-voltages.
GARFIELD++, developed for the detailed simulation of particle detectors based on ionisation measurement in gases, had already been benchmarked against the R2D2 data in a spherical geometry~\cite{Bouet:2020lbp}, and showed very good agreement.
Data were taken at different pressures between 1 and 10~bar with small high-voltages, ensuring a high drift time and no avalanche near the anode. The values of the high-voltage used in this study, which also enhance the electron attachment phenomena, are typically much lower than those used in standard detector operation conditions described in Sec.~\ref{Sec:DA}. This explains the large difference in the electron drift time (from a starting radial position at the cathode level), varying from a few ms in this study to 80/120~$\mu$s in argon/xenon, respectively, in standard detector operation.
For each pressure the signal integral ({\it i.e.} the reconstructed charge) as a function of drift time was fitted with Eq.~\ref{eq:1} to extract the electron lifetime. The results obtained at 1, 3, 6 and 10~bar in argon, with a drift time estimated using the 2023 release of GARFIELD++, are shown in Fig.~\ref{fig:PurityPressure}. Similarly the results for xenon at 1, 3 and 6~bar are shown in Fig.~\ref{fig:PurityPressureXe}.
Nonetheless, comparing different GARFIELD++ distributions, large discrepancies on the drift time were observed at small electric field. In particular, the distribution available in 2021 was benchmarked against the one released in 2023. A summary of the obtained results can be seen in Fig.~\ref{fig:PurityVsPressure}. The discrepancies between the different GARFIELD++ releases are important at pressures below $\sim 3$~bar and almost disappear at pressures above 5~bar. 

An electron lifetime below 1~ms is found in argon at 10~bar, and below 2~ms in xenon at 6~bar, showing that the gas purity does not match the quality to that already achieved by liquid noble gases experiments with electrons lifetimes greater than 20~ms in argon~\cite{Baibussinov:2009gs} and 10~ms in xenon~\cite{Plante:2022khm}.
Present performances of the R2D2 setup could certainly be improved by means of a qualitative leap in gas purity. Two identified possibilities are: baking the full device at a higher temperature, and the replacement of the recirculation pump (an important source of electronegative impurities) with a cleaner one {\it e.g.} a magnetically driven piston pump~\cite{LePort:2011hy}.


\section{Temperature effects}
\label{Sec:temp}

\begin{figure}[t]
    \centering
    \includegraphics[width=\columnwidth]{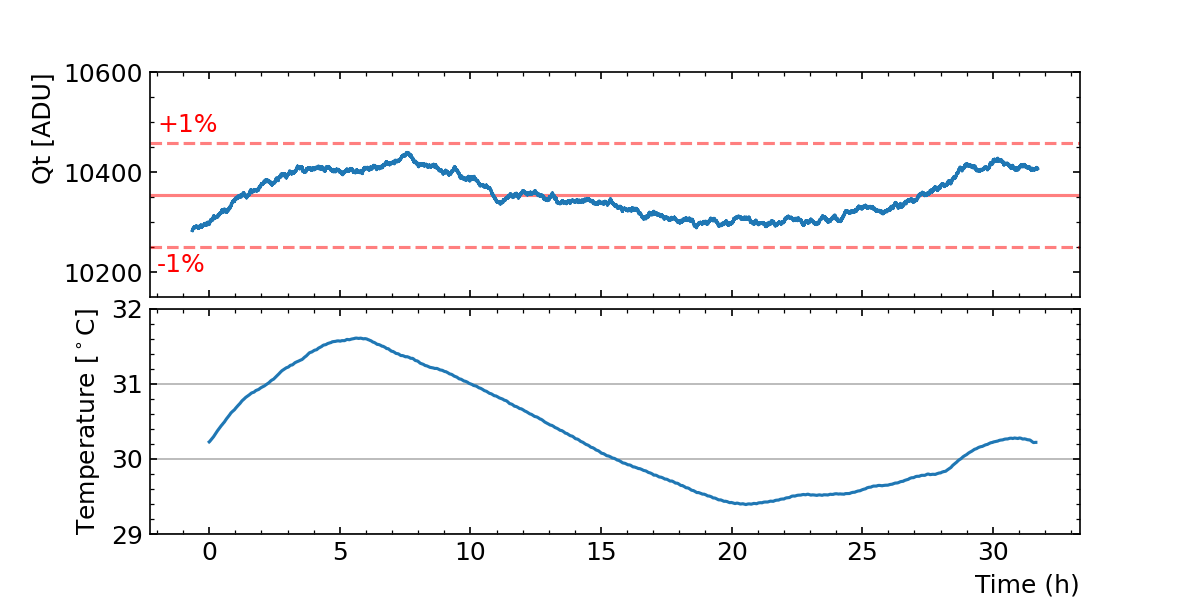}
    \caption{5.3~MeV $\alpha$ signa integral (top) and temperature on the detector surface (bottom) as a function of time.}
    \label{fig:Temperature}
\end{figure}

To precisely evaluate the energy resolution of the detector, the impact of some external factors on the detector stability were also studied. In particular, measurements conducted during the R\&D phase revealed that the electronic response is sensitive to temperature changes. Furthermore, it is plausible that temperature fluctuations in the detector's surrounding environment directly influence the internal gas pressure, thereby contributing to the observed behavior.
These effects are negligible in short runs exploiting the $^{210}$Po calibration source, since the $\alpha$ rate of few Hz allows for a large statistics in less than one hour of data taking. However, temperature changes become relevant in long runs, needed to exploit $^{222}$Rn in the detector, which last more than one day. Dedicated measurements taken while monitoring the detector temperature with a precision of 0.05 degrees Celsius, showed that the $\alpha$ signal integral variations are in phase with temperature changes. Temperature variations of about 1 degree could result in energy variations at the per-cent level, as can be seen in Fig.~\ref{fig:Temperature}.

The operation of the detector in a temperature controlled environment is desirable, however not possible for the current experimental setup. Therefore it was decided to constantly monitor the detector temperature and, as explained in Sec.~\ref{Sec:DA}, apply corrections for runs longer than few hours if needed ({\it i.e.} if temperature changes larger than 0.5 degrees were registered).

\section{Signal modelling and processing}
\label{Sec:sigp}

\subsection{Signal modelling}

\begin{figure}[t]
    \centering
    \includegraphics[width=\columnwidth]{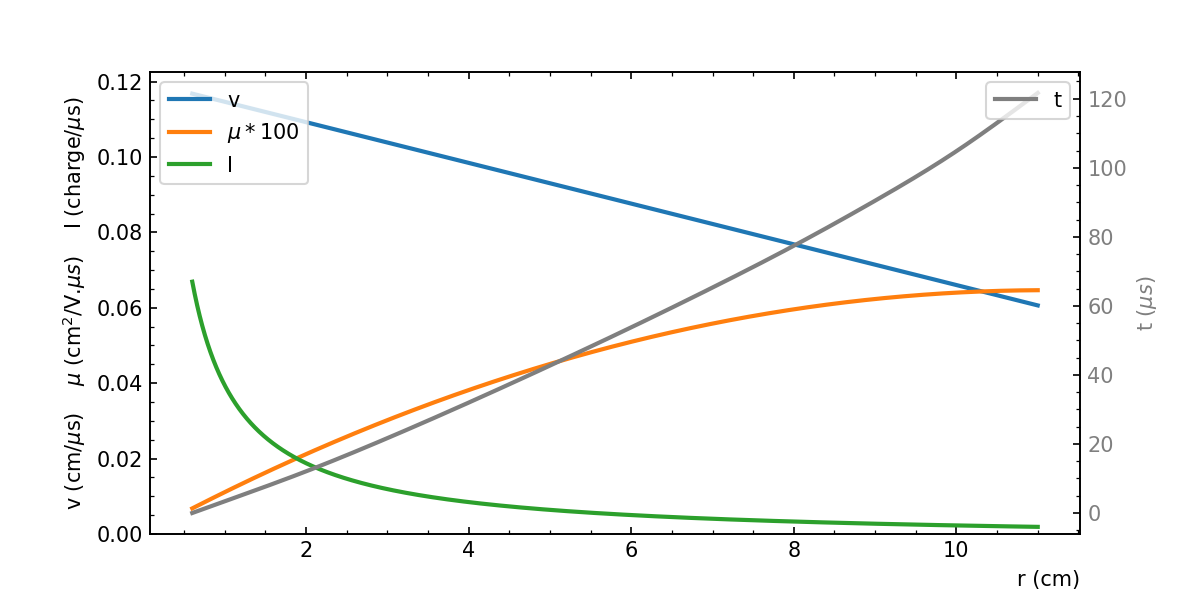}
    \caption{ Evolution of the velocity of the primaries $v$ (blue line) and the induced current $I$ (green line) as a 
	function of the distance $r$ from the anode. For information purpose, the deduced mobility $\mu$ is also shown (orange line). The 
	shown curves, built starting from the GARFIELD++ drift time simulation ($t$ shown by the grey line), were 
	computed for the CPC filled with a Xe gas at 3~bar and a high-voltage of 3000~V.}
    \label{fig:Current}
\end{figure}

\begin{figure}[t]
    \centering
    \includegraphics[width=\columnwidth]{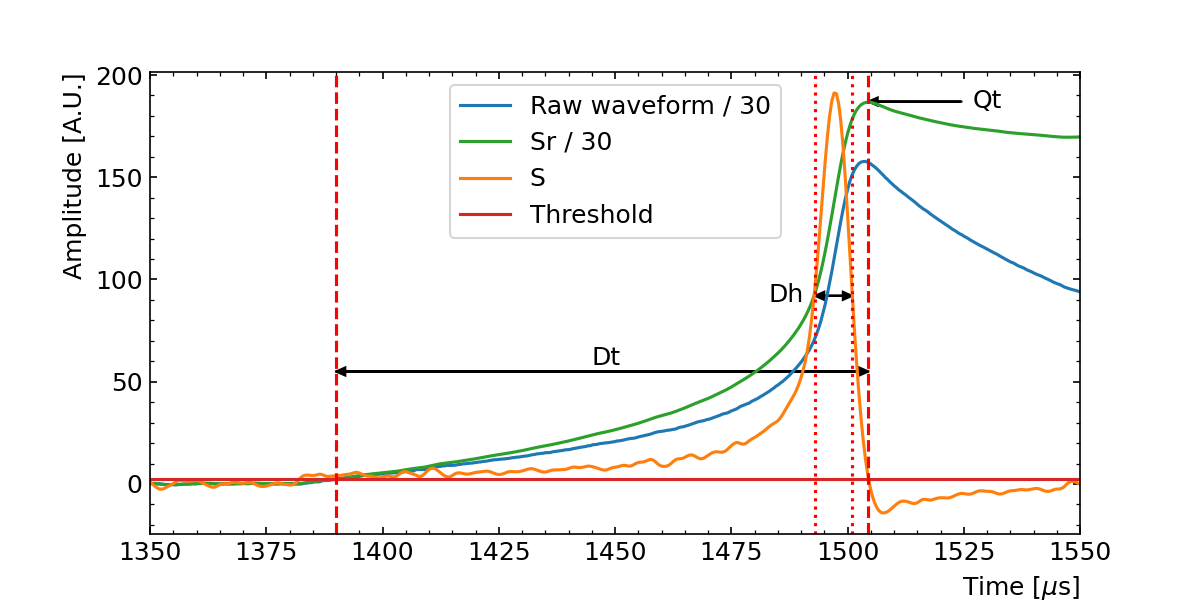}
    \caption{Example of a raw waveform (blue line), its deconvolved signal $S$ (orange line) and the reintegrated signal $Sr$ (green line). Note the scale factor of 30 for graphical purpose applied to the raw waveform and on the $Sr$ signal. The horizontal solid red line represents the threshold whereas the vertical lines indicates the bounds for the signal total width $Dt$ (dashed lines) and for the width at half maximum $Dh$ (dotted line).}
    \label{fig:dec}
\end{figure}

\begin{figure*}[t]
    \centering     
   	 \subfigure[\label{fig:QtDtvsr}]{\includegraphics[width=\columnwidth]{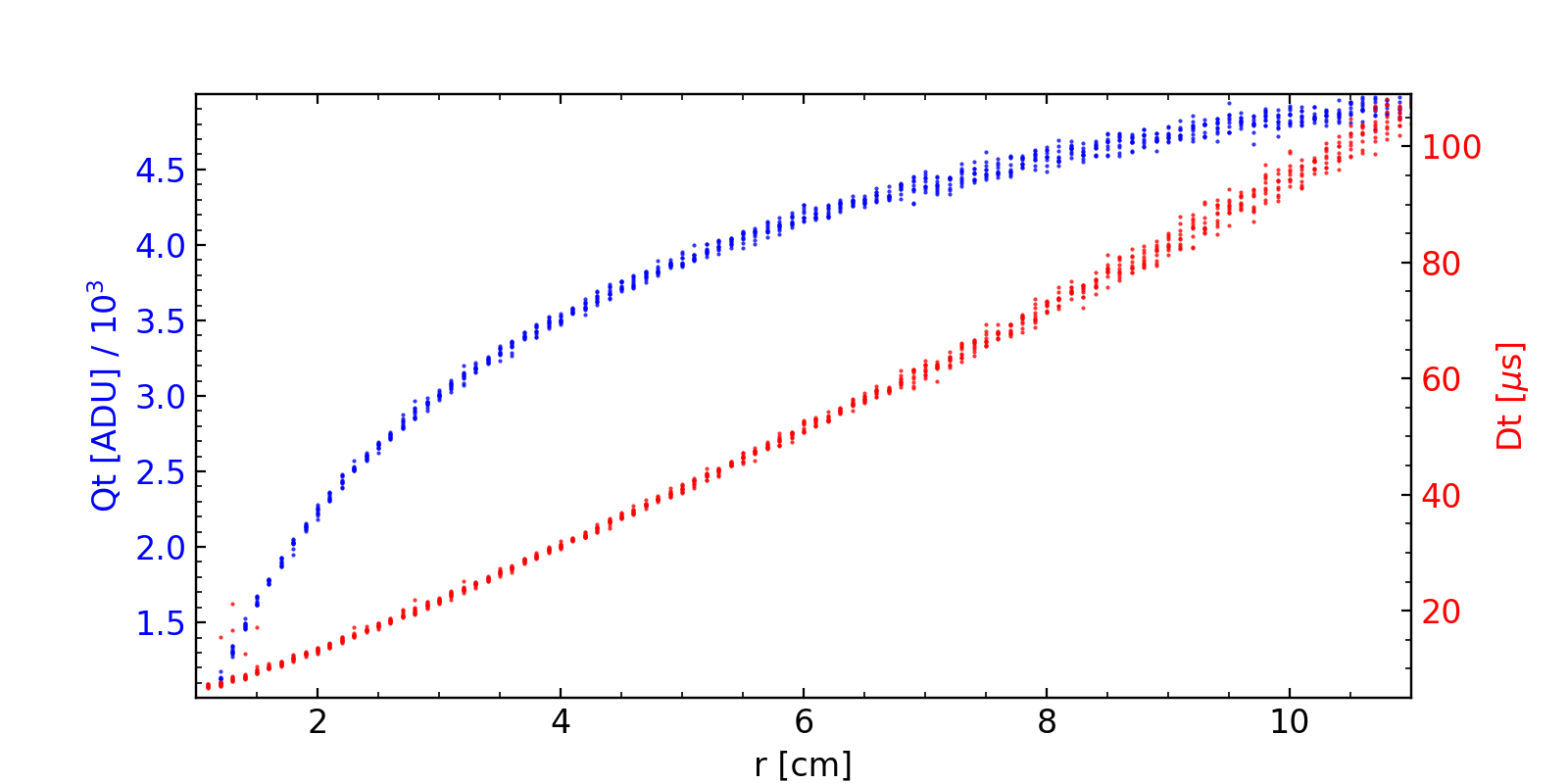}}
    	 \subfigure[\label{fig:dtvsqt}]{\includegraphics[width=\columnwidth]{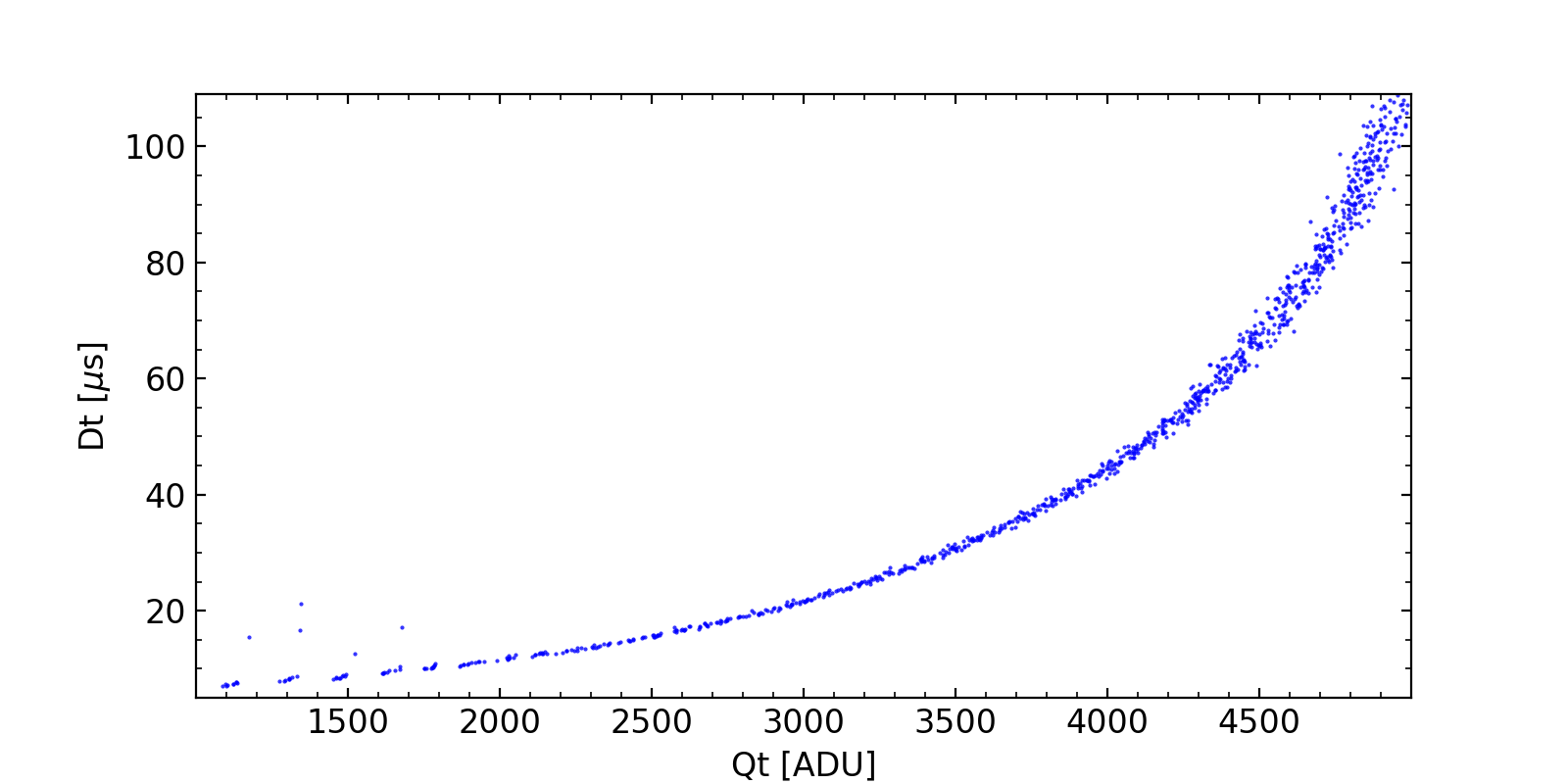}}
      	 \subfigure[\label{fig:dhvslength}]{\includegraphics[width=\columnwidth]{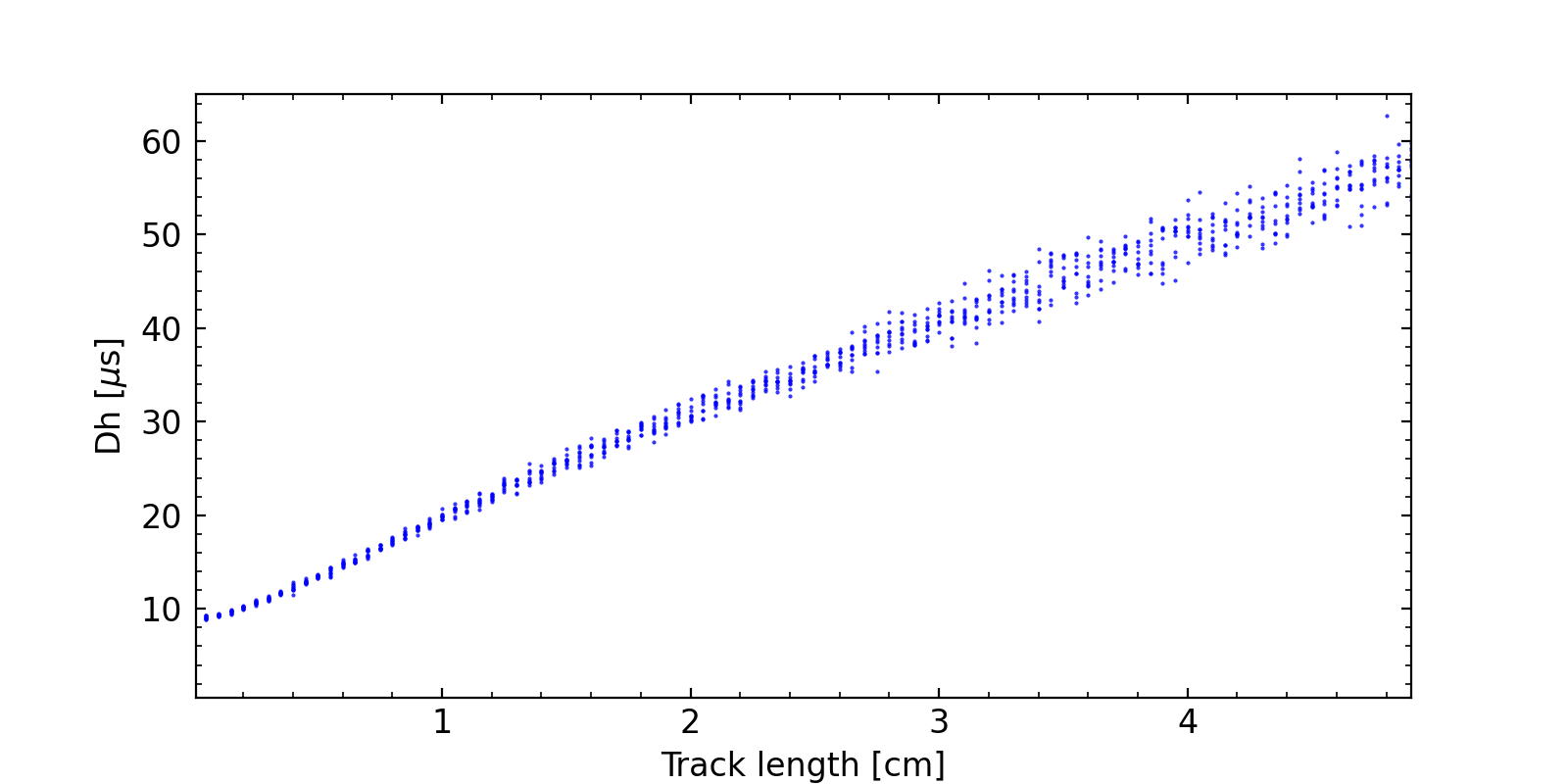}}
    \caption{Signal modeling results. $Qt$ (blue) and $Dt$ (red) as a function of the distance $r$ to the anode for a point-like energy deposit (\ref{fig:QtDtvsr}). $Dt$ versus $Qt$ (\ref{fig:dtvsqt}).
    $Dh$ versus the radial extension of the track, for tracks emerging from the cathode and heading towards the anode (\ref{fig:dhvslength}).}
    \label{fig:DtDhQt}
\end{figure*}

The principle of calculating the current induced by the movement of charges between the electrodes 
has been established for decades~\cite{Ramo:1939vr,Shockley:1938itm}. This method is widely used in the field of particle 
detection simulation~\cite{Dris:2014qpa}. In a parallel plate ionization chamber (PPC), examining the current signal enables the extraction of various physical information related to the interaction. This includes details such as the deposited energy and the distance of the interaction from the anode~\cite{Recine:2014zca,Warburton:2004nql}. 
These features led to the selected approach based on the exploitation of the current signal. 
Such a signal is extracted by digital 
processing of the original output waveform of the integrator~\cite{Bouet:2020lbp,Bouet:2023zyk,Lautridou:2023vrk}.
To interpret the results, a dedicated modeling of the expected current waveform was jointly developed.
Compared to GARFIELD++, it is a simplified and fast simulation based on the average macroscopic behavior of the primaries 
during their transport in the gas. This choice was motivated by two observations: first the main parameters which shape the signal are more easily identifiable; and secondly, in the ionization regime (where the signals are very weak) the fluctuations observed are essentially dominated by Radio Frequency Interferences (RFIs) and electronic noises. 
Starting from the GARFIELD++ results giving the drift time of a primary as a function of the distance, the average velocity of the charge $v_e(r)$ is first deduced. The expression of the electric field of a cylindrical counter is given by:
 \begin{equation}
 \label{eq:2}
E(r)  = V_0 \times E_w(r) = V_0 \times\frac{1}{r} \times \frac{1}{log(r_{cathode}/r_{anode})},
\end{equation}
 where $r$ is the radial distance from the anode, $V_0$ the potential difference between anode and cathode, $E_w$ the so-called weighted electric field, and $r_{cathode}/r_{anode}$ the ratio of cathode and anode radii, respectively.
By taking into account Eq.~\ref{eq:2} and referring to the Shockley-Ramo theorem, for this simple geometry the current $I(r)$ induced on the anode by an elementary charge drifting all along the radius of the CPC  can be calculated via:
 \begin{equation}
 \label{eq:3}
I(r)  = e \times  v_e(r) \times E_w(r), 
\end{equation}
where $e$ is the electron charge and $v_e(r)$ the drift radial velocity at the $r$ position.
An example of the shape of this current is presented in Fig.~\ref{fig:Current} for the CPC filled with a Xe gas at 3~bar and a high-voltage of 3000~V.
Note that in the developed simulation, the contribution of ions is omitted because their speed is several 
orders of magnitude lower (while in proportional mode, most of the signal is induced by the ions 
produced in the avalanche). Furthermore, the very long ion component would be well below the detector sensitivity, particularly in xenon.
The introduction of the distribution of primary charges and the conversion of the distance into sampled time then allow the construction of the signal induced by a full track.
The penultimate stage of the simulation chain introduces the attachment effect as well as the instrumental noise.
Convolution of the resulting waveform with the electronics response completes 
the construction of the induced current signal, called $S(t)$. 
While constant for a PPC, in case of the CPC, the $S(t)$ signal varies considerably in time (see Fig.~\ref{fig:dec}), usually increasing slowly at first.
Consequently, the conventional observables used for PPCs are no longer suitable, and new physical observables must be constructed to correctly classify the registered events.
This necessitates first fixing the analyzed relative fractions of the signal. Since $S(t)$ is typically small at its onset, the ideally reintegrated waveform $Sr(t)$ is employed for this purpose.
Indeed, the latter makes it possible to immediately deduce the total induced charge $Qt$, which is given by the maximum of $Sr$ (the integral of $S$). 
This allowed us, for temporal estimation, to precisely set the crossing points above the thresholds, defined in terms of a fraction of $Qt$.
With both temporal representations, it becomes possible to build the three observables used in the analysis, namely $Qt$, $Dt$ and $Dh$,  which contain different information about the interaction in the gas. 
An example of a signal with the associated observables is presented in Fig.~\ref{fig:dec}.
The left part of the signal is due to the induction in the weakly increasing electric field, far from the anode. In this region of electric field, the detector functions almost as a PPC. The strong growth on the right part of the signal is caused by the rapid increasingly variation of the electric field (as $1/r$) in the region $\sim 1$~cm from the anode~\cite{Sauli77}. This is typical of a signal observed with a wire proportional chamber (WPC)~\cite{Schwegler2014}.  
After reaching a maximum, as the primaries are collected by the anode, 
the induction vanishes. The descending part of the transient becomes mostly governed by the impulse response of the amplifier (undershoot), rather than by the arrival tail of the electrons.
Each observable describes a specific feature of the registered signal:
\begin{itemize}
\item $Qt$ is the maximum value of the integrated signal $Sr$, noting that 
because of the non constant electric field, it is not directly the true initial ionization charge $Q$.
To reconstruct the true energy deposited (at first order proportional 
to the number of primary charges created), $Qt$ must be corrected for the radial position of the interaction, especially when the interaction is close to the anode. 
However, by selecting a narrow range in energy deposit positions, $Qt$ can provide a rough estimate of the deposited energy.
To complete the specificity of the ionization signal as a consequence of the Shockley-Ramo theorem, it should be also noted that $Qt$ is not affected by diffusion during the transport of primaries.
\item $Dt$ is the total duration of the signal $S$ from the moment it rises above threshold (constant fraction of the total integral) up to the maximum of the signal $Sr$ ({\it i.e.} when $S$ goes to zero).  
As is the case in a PPC, $Dt$ gives a direct measurement of the maximum radial distance from the track to the anode (see Figs.~\ref{fig:QtDtvsr} and~\ref{fig:dtvsqt}).
\item $Dh$ is the width of the signal $S$ at half height. This observable has no specific meaning in a PPC, and is linked to the radial extent of the track demonstrated by the developed signal simulation for the CPC.
This feature makes $Dh$ particularly interesting in the event selection applied for the energy resolution analysis. 
From the Shockley-Ramo theorem it follows that for a point-like energy deposit, $Dh$ is independent on the radial position inside the CPC. However, a track can be considered as the sum of energy deposits at the same time and different radial distances. Each one would be the seat of the same radial evolution, but shifted by the difference in drift time at the considered positions. Such a superposition results in a signal with a larger $Dh$ for tracks extending in the radial direction (see Fig.~\ref{fig:dhvslength}). 
\end{itemize}
As an example of the behavior of these quantities, the current waveforms, including noise and electronic response, were modeled in 1~mm steps from the cathode to 1~cm from the anode. Predictions of the observables $Qt$ and $Dt$ as a function of the distance $r$ to the anode for a point-like deposit are presented in Fig.~\ref{fig:QtDtvsr}. The dependencies demonstrate that an effective recognition of the endogenous decays (betas and alphas) emitted by the anode or the cathode can be obtained, making it possible to better reject this contribution from the radioactive background. In Fig.~\ref{fig:dtvsqt} the dependence of $Dt$ as a function of $Qt$ is shown, whereas
a simulation of the evolution of $Dh$ as a function of the track length is presented in Fig.~\ref{fig:dhvslength}. Ionization lines starting from the cathode and pointing towards the anode were selected. For inclined tracks, projecting the tracks onto the radial axis ultimately results in a similar distribution being followed.

The computed response of the CPC in ionization mode combines advantages specific to PPC with some distinctive of WPC. The CPC signal is produced throughout the drift of the electrons towards the anode, offering original and attractive detection functionalities, in particular with a single read-out anode.

\subsection{Signal processing}

\begin{figure}[b]
    \centering
    \includegraphics[width=\columnwidth]{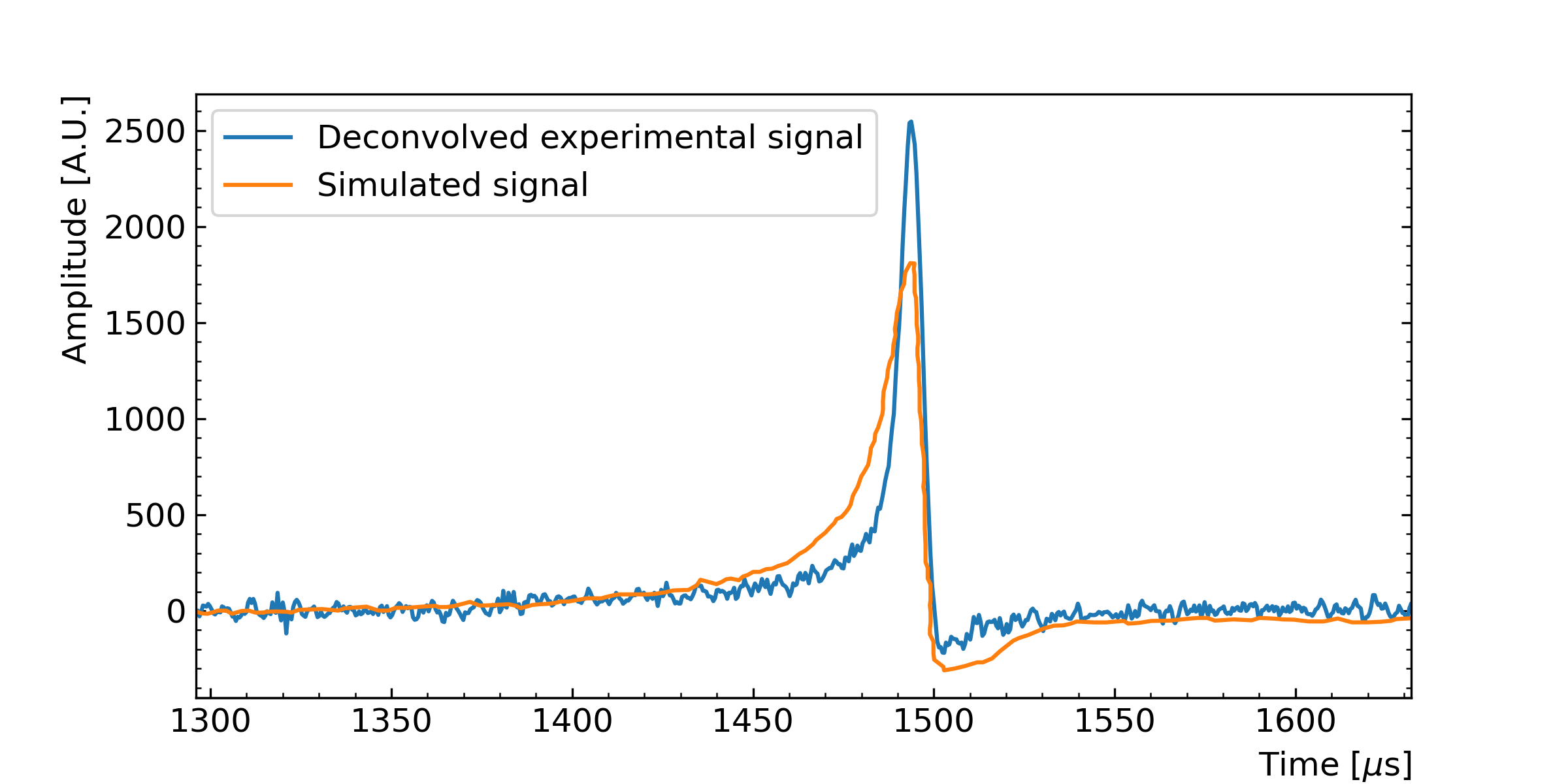}
    \caption{Experimental waveform after processing for an $\alpha$ particle issued	from $^{210}$Po (blue) compared with a simulated signal due to electrons	created at the cathode radial position (orange).}
    \label{fig:SigSim}
\end{figure}

Regarding the experimental data, which relies on the treatment 
of the registered integrator waveform, several numerical operation 
are needed to retrieve the current signal. 
The first step consists of the application of a Fourier transform base filter with the aim of mitigating the noise caused by the electronics and the detector vibrations. A deconvolution algorithm is then applied in order to retrieve the current signal. 
The registered waveform is indeed a convolution of the physical signal with the typical RC shaping of the preamplifier: such a component has to be removed to fully exploit the temporal
shape and correctly infer physical properties of the event. 
A last filtering step in temporal space serves to mitigate high frequency noise brought by the deconvolution algorithm. An ultimate adjustment of the baseline completes the processing, leading to the final deconvolved time series $S$. 
Like for the simulation, a running sum of $S$ provides the signal $Sr$.  
The comparison of a registered waveform corresponding to an $\alpha$ particle issued from $^{210}$Po, with the simulation of the signal due to ionisation electrons created at the cathode surface, can be seen in Fig.~\ref{fig:SigSim}. The small difference in shape was attributed to the slight deviation of the average distribution of migration times from GARFIELD++ with respect to the real behavior of the charges in the detector.

\begin{figure*}[tp]
    \centering     
   	 \subfigure[\label{fig:ExampleBiplot1}]{\includegraphics[width=\columnwidth]{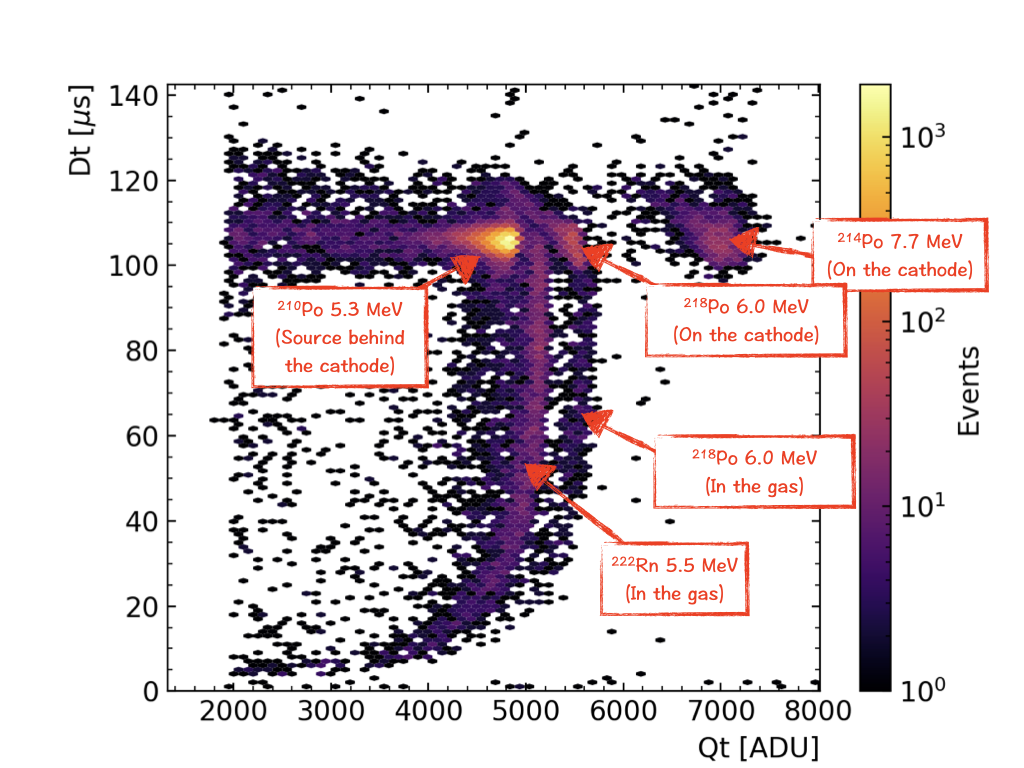}}
    	 \subfigure[\label{fig:ExampleBiplot2}]{\includegraphics[width=\columnwidth]{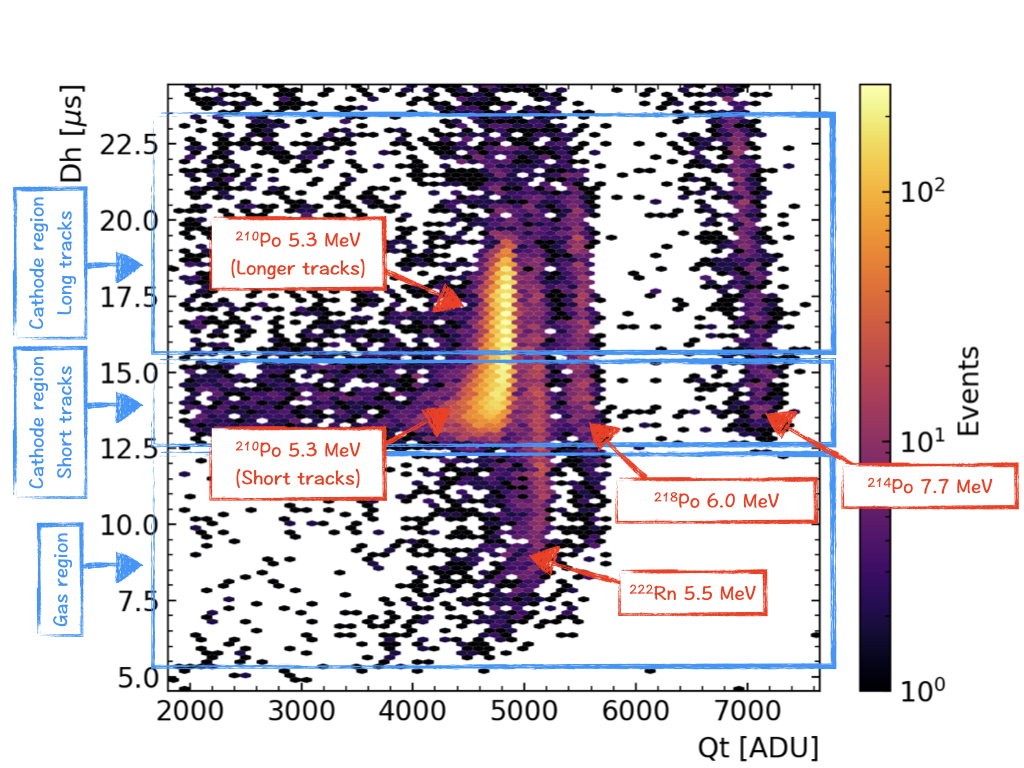}}
    \caption{Example of a bi-plot $Dt$ versus $Qt$ (\ref{fig:ExampleBiplot1}), and  $Dh$ versus $Qt$ (\ref{fig:ExampleBiplot2}) for a run with the detector filled with xenon at 3~bar, a central anode of 1.2~mm, and a high-voltage of -3000V applied to the cathode.}
    \label{fig:ExampleBiplot}
\end{figure*}

 To graphically show the different event populations, an example corresponding to a run taken with the detector filled with xenon at 3~bar is shown in Fig.~\ref{fig:ExampleBiplot}. The different sources ($^{210}$Po calibration source and $^{222}$Rn contamination) are shown as well as their expected position. 
 In the $Dh$ versus $Qt$ plot, the $^{210}$Po structure is explained considering that 5.3~MeV $\alpha$ tracks are short ({\it i.e.} 3-4~mm) and they traverse a small hole in the cathode to reach the active volume. For such short tracks the hole thickness of 200~$\mu$m is not negligible and in such a region the field is distorted, therefore not all ionization electrons will reach the central anode. This explains why longer tracks inside the volume ({\it i.e.} larger values of $Dh$) have a better energy resolution, whereas shorter tracks, partly absorbed in the cathode, exhibit a tail towards lower values of the reconstructed energy $Qt$.

\section{Data analysis}
\label{Sec:DA}

\subsection{Data-taking conditions}

 The primary goal of the R2D2 R\&D is to establish the detector capability, when filled with xenon gas, of reaching percent-level energy resolution. A second objective is the demonstration that the same energy resolution can be achieved with the same detector filled either with xenon or argon. This would be an important asset for the future of the project allowing the validation of larger prototypes with argon, in a much more economic way, and with no need of a gas recuperation system.
 As mentioned in Sec.~\ref{Sec:intro}, to reach the project goals it was decided to run the detector in ionisation mode. Such a choice is driven by several physical and practical reasons argued below.
 \begin{itemize}
\item Provided that the signal-to-noise ratio is sufficient, operation in ionisation mode inherently yields the best possible energy resolution avoiding any stochastic fluctuation on the collected charge due to the avalanche process. Furthermore, tiny inhomogeneities of the wire diameter have no impact in ionisation mode, whereas in proportional mode they result in different gains as a function of the position along the wire impacting the energy resolution.
\item In the ionisation regime the signal is mostly given by the electrons drift which is similar between xenon and argon: the similar electron mobility allows for drift times of the order of 120~$\mu$s in xenon and 80~$\mu$s in argon. On the other hand, the proportional signal is mainly due to the drift of ions produced in the avalanche. With about three orders of magnitude smaller ion mobility in xenon, the time scales become very different for the two gases, and electronegative impurities are much more critical in xenon.
\item To work in proportional at high pressure a very thin wire or high voltages are needed. In the current setup the high-voltage power supply  as well as the connections are validated up to 6~kV only. In addition when such a high bias is applied to the cathode a small leak current is observed, as well as an increase of noise. To overcome this issue several modifications would have been necessary, such as increasing the distance between the cathode and the grounded sphere, and replacing most of the connections. A fine wire could nevertheless be used but at the cost of reducing the electric field at the cathode, making the measurements more sensitive to electronegative impurities of the gas. As explained in Sec.~\ref{Sec:gaspur}, the gas purity of the R2D2 setup is an important parameter and a weaker electric field would be unsustainable to operate without problems. 
\end{itemize}

Data taken with two different anode thicknesses (50~$\mu$m and 1.2~mm diameter) were compared. The advantage of a thick anode lies in the ability to remain in the ionization regime and to increase the high-voltage (\textit{i.e.}, the electric field), thereby reducing the drift time and mitigating the impact of electronegative impurities. This is not the case for a thin wire since an increase of the high voltage could initiate an avalanche, bringing the detector into the proportional regime. For this reason the detector operation in ionisation mode at 6~bar with xenon was possible only with the 1.2~mm thick central anode.

Concerning the gas pressure, a scan between 1 and 10~bar was performed in argon, where the upper limit is given by the maximal operating pressure of the hot getter. The runs with xenon inside the detector were limited to 6~bar by the maximal volume allowed by the gas recovery system.

\subsection{Event selection}
To evaluate the energy resolution, two samples were selected, corresponding to $\alpha$'s of 5.3~MeV issued by $^{210}$Po decays and $\alpha$'s of 5.5~MeV issued by $^{222}$Rn decays. The first sample comes from the calibration source located behind the cathode as explained in Sec.~\ref{Sec:exp}. Such a sample has the advantage of being well localized with a known starting position of the $\alpha$ tracks. The second sample comes from radon which is introduced into the detector because of the recirculation through the cold getter. This is a well known issue of the cold getter for radio-pure experiments, but one that has turned into an advantage in providing a diffuse source for current testing. This decay allows the verification that the energy resolution is homogeneous throughout the detector volume, although is impacted by edge effects not corrected in the current setup ({\it e.g.} distortion of the field lines towards the end-caps).

\begin{figure}[t]
    \centering
    \includegraphics[width=\columnwidth]{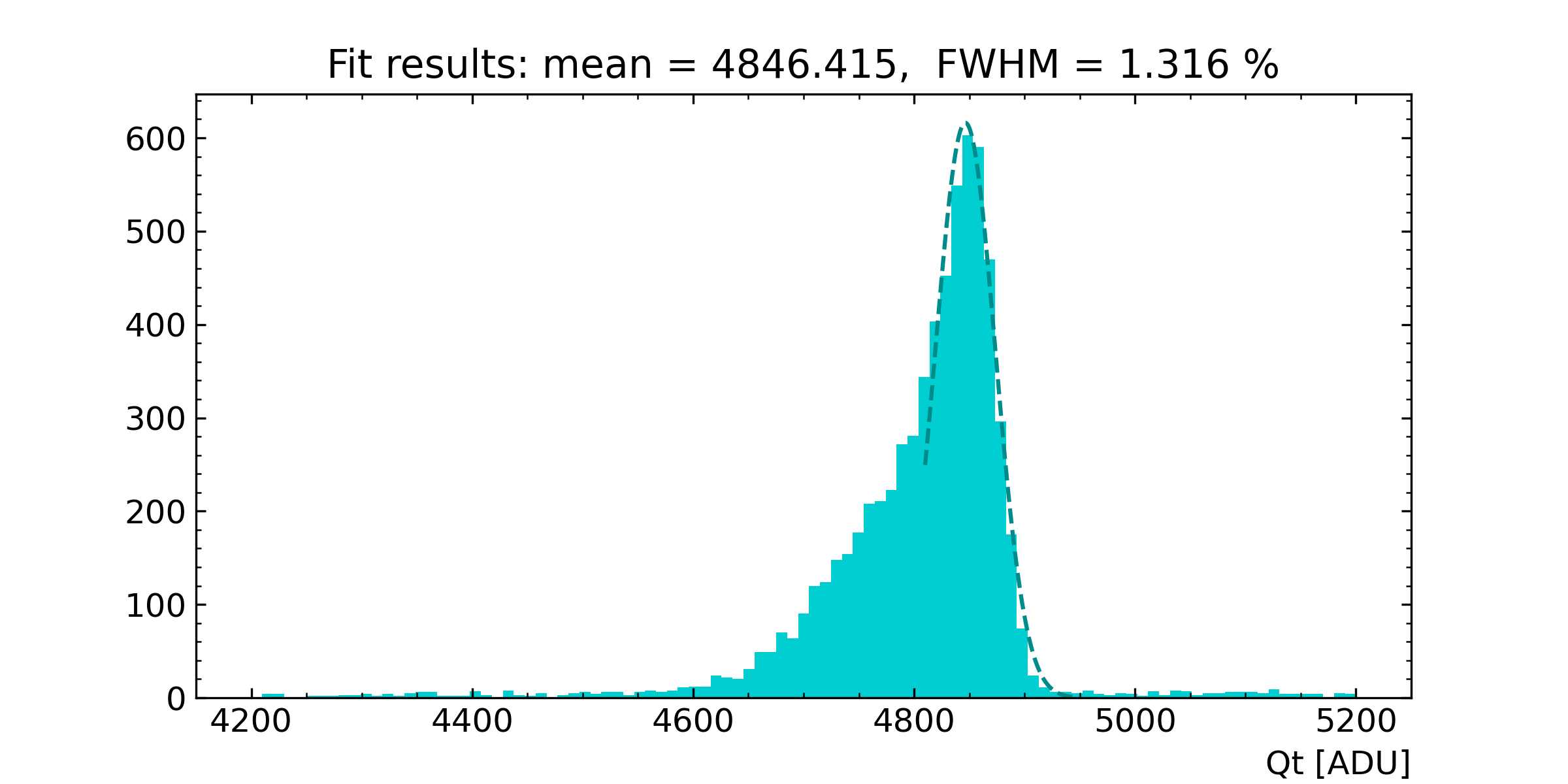}
    \caption{Resolution fit for 5.3~MeV $\alpha$s from $^{210}$Po source for a run with the detector filled with xenon at 3~bar, an anode of 1.2~mm, and a high-voltage of -3000V applied on the cathode. A selection of the events with $100 \mu s < Dt <110 \mu s$ and $Dh >15 \mu$s has been applied.}
    \label{fig:ResEx}
\end{figure}

The $^{210}$Po events are selected in a $Dt$ interval (dependent on the gas pressure and on the high-voltage applied) corresponding to events originating at the cathode radial position. Events rejected with a $Dt$ below the lower accepted bound correspond to events originating in the gas volume closer to the anode, whereas events rejected (with $Dt$ above the upper bound) correspond to pile-up events: those in coincidence with cosmics or low energy gammas. The selection is tuned for each run according to the gas and the high-voltage: for the run with the detector filled with xenon at 3~bar, the anode of 1.2~mm diameter and a high-voltage of -3000V (see Fig.~\ref{fig:ExampleBiplot1}), events with a $Dt$ between 100 and 110~$\mu$s are selected.
A second selection for $^{210}$ Po events is based on $Dh$, which, as shown in Fig.~\ref{fig:dhvslength},  is an observable related to the radial length of the track. To minimize the impact of the energy loss in the hole through which $\alpha$ particles traverse the cathode, feature specific to the source location in the current setup, longer $\alpha$ tracks inside the detector are selected. This is done applying a lower cut on $Dh$ in the data selection for the resolution analysis.
Even in this case the cut depends on the run condition. For 3~bar of Xe (see Fig.~\ref{fig:ExampleBiplot2}) events with $Dh$ above 15~$\mu$s were selected. For the latter, the reconstructed energy is fitted with a Gaussian function that starts approximately halfway up the low energy side and continues until the end of the peak on the high energy side. All $\alpha$'s tracks pass through the cathode hole, but despite the selection cutoff $Dh$, a low energy tail due to partial energy loss persists in the selected data. An example is shown in Fig.~\ref{fig:ResEx}.

 \begin{figure}[t]
    \centering
    \includegraphics[width=\columnwidth]{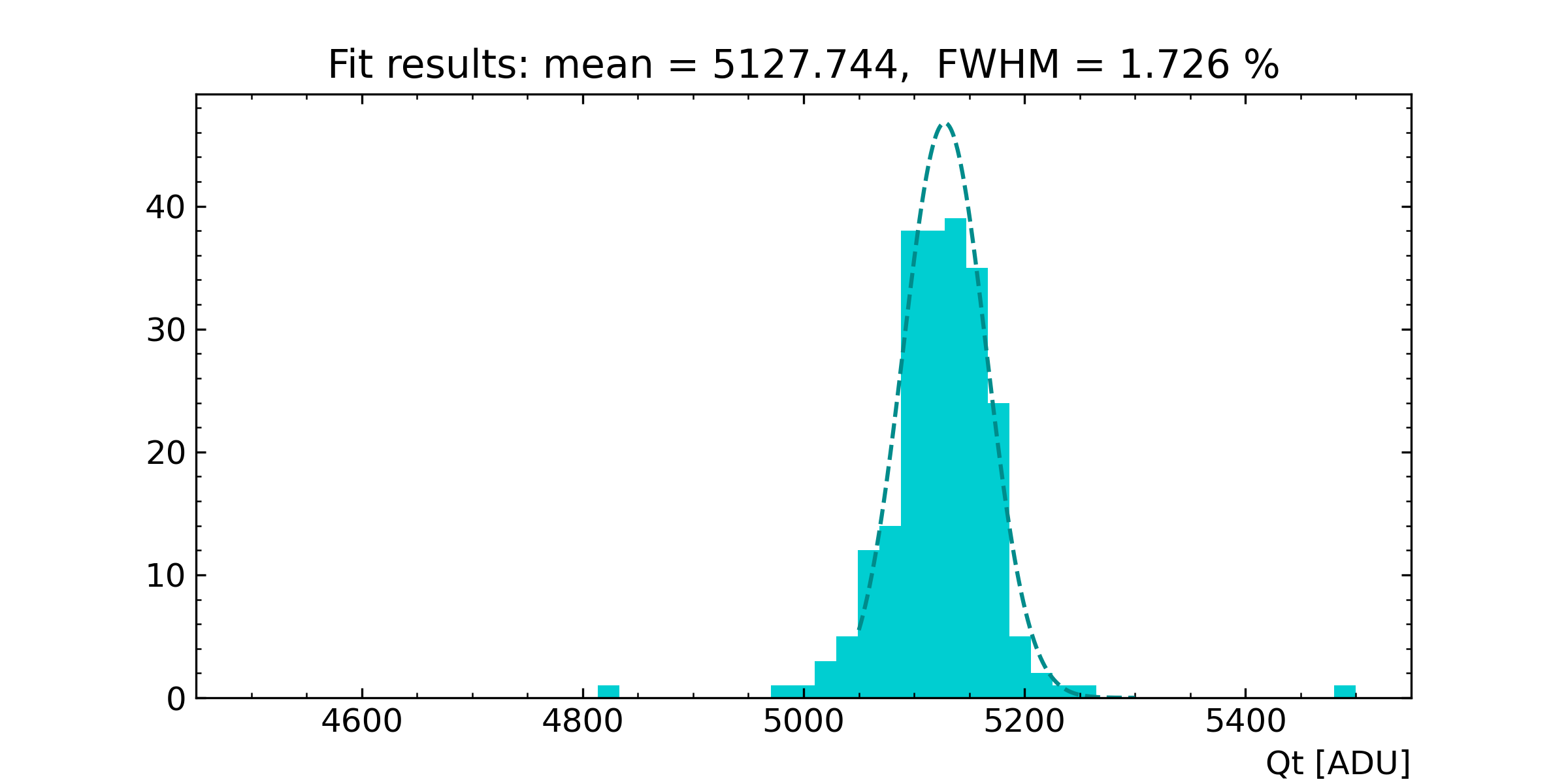}
    \caption{Resolution fit for 5.5~MeV $\alpha$s from $^{222}$Rn source for a run with the detector filled with xenon at 3~bar, an anode of 1.2~mm, and a high-voltage of -3000V applied on the cathode. A selection of the events with $70 \mu s < Dt <100 \mu s$ and $Dh <12.5 \mu$s has been applied, in addition to the temperature and position corrections}
    \label{fig:ResExRn}
\end{figure}

\begin{figure*}[t]
    \centering     
   	 \subfigure[\label{fig:ResPo}]{\includegraphics[width=\columnwidth]{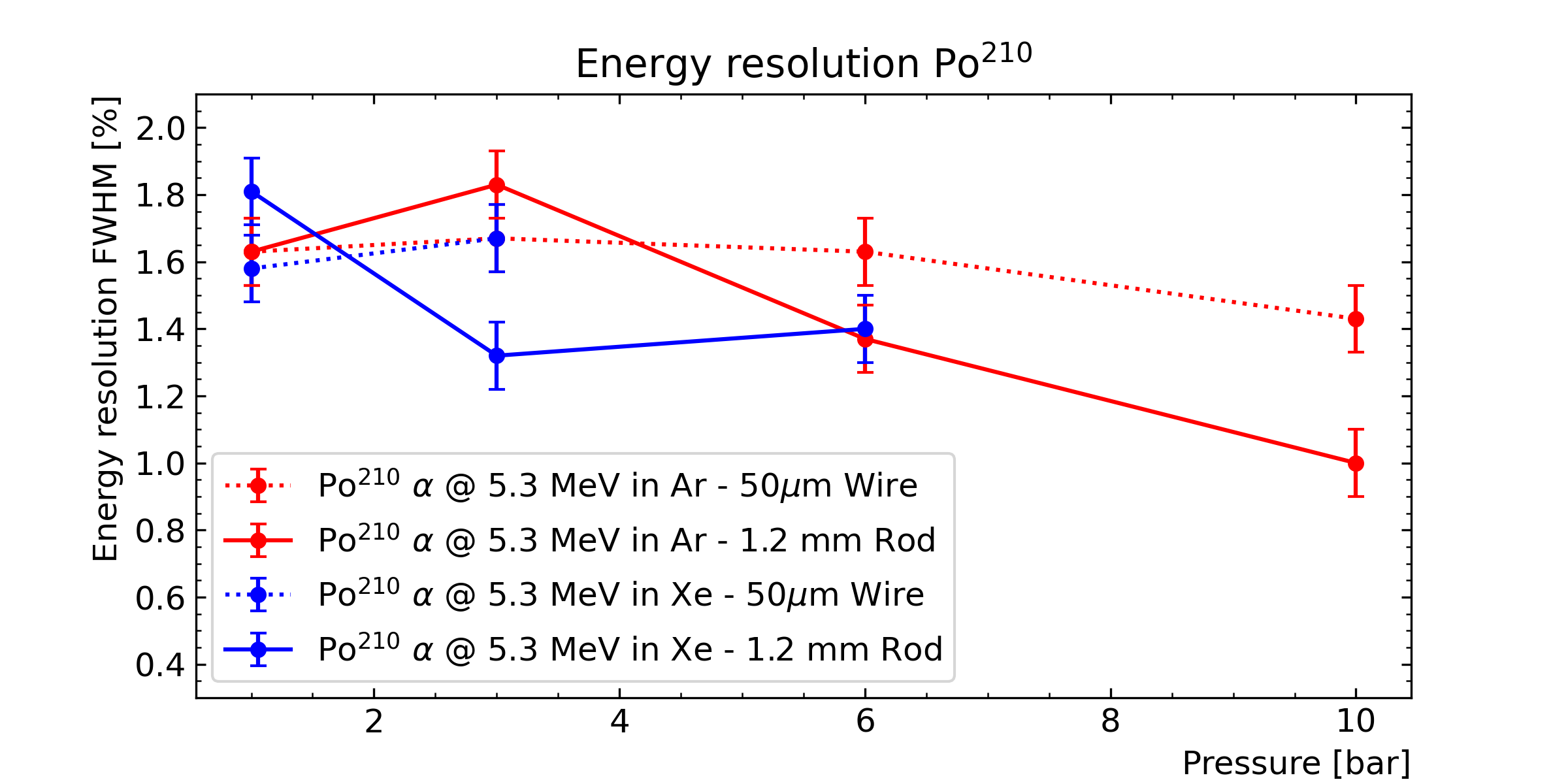}}
    	 \subfigure[\label{fig:ResRn}]{\includegraphics[width=\columnwidth]{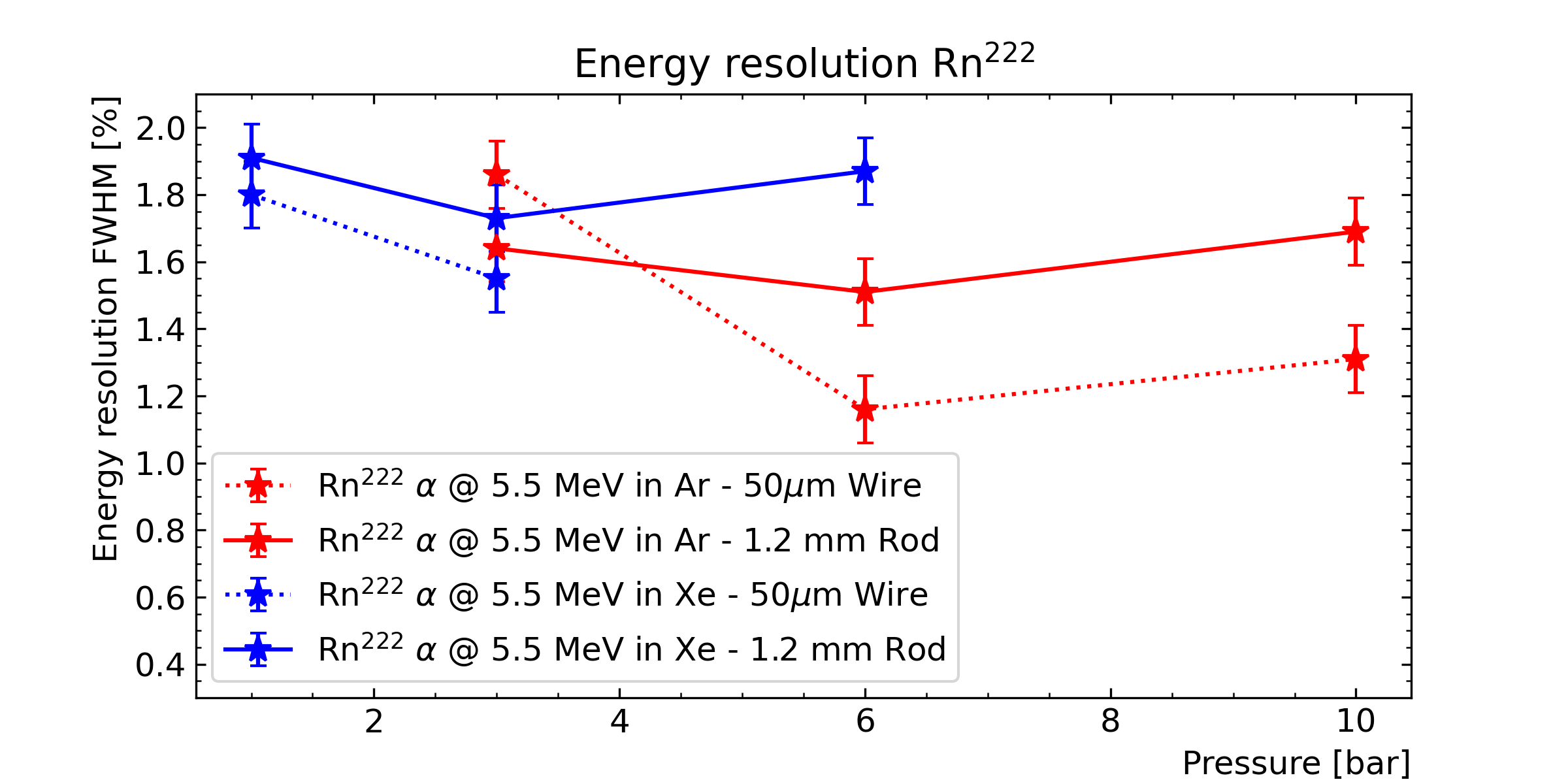}}
    \caption{Energy resolution obtained with 5.3~MeV $\alpha$''s from $^{210}$Po (\ref{fig:ResPo}), and 5.5~MeV $\alpha$'s from $^{222}$Rn (\ref{fig:ResRn}). The blue lines represent xenon measurements whereas the red lines stands for argon measurements. Dotted lines are used for 50~$\mu$m anode whereas solid lines are used for 1.2~mm anode.}
    \label{fig:Res1}
\end{figure*}

\begin{table*}[ht]
\begin{center}
\begin{tabular}{|c|c|c|c|c|c|c|}
\hline
Gas& Pressure (bar) & HV (V) & Gain & Anode diameter & Resolution $^{210}$Po &  Resolution $^{222}$Rn \\
\hline
\multirow{5}{*}{Xe}& 1 & 700 & 1 &\multirow{2}{*}{50~$\mu$m}  & 1.6\% &   1.8\% \\
& 3 & 1800 & 1.8 &  & 1.7\% &   1.6\% \\
\cline{2-7}
& 1 & 1000 & 1 & \multirow{3}{*}{1.2~mm} & 1.8\% &   1.9\% \\
& 3 & 3000& 1 & & 1.4\% &   1.7\% \\
& 6 & 5500 & 1 &  & 1.4\% &   1.9\% \\
\hline
\multirow{8}{*}{Ar}& 1 & 700 & 1.3 &\multirow{4}{*}{50~$\mu$m}  & 1.6\% &   - \\
& 3 & 200& 1 & & 1.7\% &   1.9\% \\
& 6 & 1700& 1.2 & & 1.6\% &   1.1\% \\
& 10 & 2200& 1.1 & & 1.4\% &   1.3\% \\
\cline{2-7}
& 1 & 500 & 1 & \multirow{3}{*}{1.2~mm} & 1.6\% &   - \\
& 3 & 1000& 1 & & 1.8\% &   1.6\% \\
& 6 & 1500& 1 & & 1.4\% &   1.5\% \\
& 10 & 3000& 1 & & 1.0\% &   1.7\% \\
\hline
\end{tabular}
\caption{Summary of all measurements taken, corresponding to Fig~\ref{fig:Res1}.}
\end{center}
\label{Tab:res}
\end{table*}%

The $^{222}$Rn event selection follows the same logic used for $^{210}$Po, with some additional treatments related to the temperature variation and the energy dependence according to the initial radial event position. Considering the radon event rate is approximately an order of magnitude lower than that of the polonium source, to collect sufficient statistics the analysis duration should be at least 12 hours.
As discussed in Sec.~\ref{Sec:temp}, variations of temperature at the level of 1 degree have an impact on the reconstructed charge at the percent level, thus degrading the energy resolution. The R2D2 setup is operated in an environment with variation of temperature between night and day which can reach up to few degrees. It follows that a correction based on the temperature variation becomes mandatory to obtain the desired energy resolution with the $^{222}$Rn. Such a correction was systematically applied  in each radon analysis.
 The second adjustment needed for radon events study aims to compensate for the position dependence of the reconstructed energy. As is seen in Fig.~\ref{fig:ExampleBiplot1}, events at small $Dt$ ({\it i.e}. close to the anode) exhibit a small reconstructed charge, as perfectly predicted by the simulation presented in Sec.~\ref{Sec:sigp}. Also, previous work showed that the dependence of proportional signals on the initial radial position could be well described by a power law~\cite{Lautridou:2023vrk}. Although not optimal here, it was assumed that this dependency might still apply because both induction principles were analogous. Thus, part of the data was used to estimate such a function, which was then applied on the remaining data to correct for this effect.
 Once temperature and position dependence corrections are accounted for, data are selected using $Dt$ and $Dh$ observables. An upper bound on $Dt$ is needed to reject all events coming from the cathode and clean the sample from tails of the $^{210}$Po source events. A lower cut on $Dt$ is also applied to reject the region close to the anode, where the position dependence on the reconstructed charge is important and uncertainties on the model used for the correction could result in a resolution degradation. For the run with the detector filled with xenon at 3~bar, the anode of 1.2~mm radius and a high-voltage of -3000V (see Fig.~\ref{fig:ExampleBiplot1}), events with a $Dt$ between 70 and 100~$\mu$s are selected.
 An upper bound on $Dh$ also helps to reject events from the cathode and clean the sample from polonium events. For the xenon run at 3~bar (see Fig.~\ref{fig:ExampleBiplot2}) events with $Dh$ below 12.5~$\mu$s are selected.
 An example of the fitted Gaussian function is shown in Fig.~\ref{fig:ResExRn}, where the low energy bound is almost symmetric to the high energy one. Indeed no source structure effect is present for radon, although a small tail of lower energy events can still be present due to detector edges effects.

\subsection{Results}

The obtained results are summarized in Fig.~\ref{fig:Res1} and Tab.~\ref{Tab:res}. The error at the level of 0.1\% was estimated varying the selection cuts and the fit range. Nonetheless the detector running conditions in terms of vibrational and electronic noise were not constant between different runs. 
In particular, part of these disturbances was attributed to the electrical environment of the experimental room, which accommodates other measuring devices. 
This uncontrolled effect is probably the largest source of systematic error in the performed measurements and it could account for variations at the level of 0.2\% (variations observed in runs taken in the same condition of gas pressure and high-voltage). 

The analysis performed on 5.3~MeV $\alpha$'s emitted by $^{210}$Po (Fig.~\ref{fig:ResPo}) shows that a resolution at the level of 1.5\% can be obtained independently on the gas pressure. Furthermore, similar results have been obtained for xenon and argon gases. This clearly suggests that the performance of future prototypes could be first established using argon, and extrapolated to xenon, at least if similar gas purities can be obtained. The thick anode seems to yield a better energy resolution, in particular at high pressure. The fact that the thicker anode produces a higher electric field at the cathode, thereby reducing the attenuation due to electron attachment during drift, could explain partly this observation.

The analysis performed on radon events confirms the independence of the energy resolution from the gas pressure and nature ({\it i.e.} argon or xenon). Switching from a well localized source on the cathode to a diffuse one does not degrade the energy resolution. These observations demonstrate that the energy resolution remains independent of the radial position of the energy deposit, despite a strongly varying electric field. 
However, a slight deviation between radon and polonium results can be observed. Unlike polonium events, the thinner wire yields a slightly better resolution for radon events at high pressures.
A possible explanation is that for radon decays, happening throughout the volume and closer to the cathode, the attachment effect is weaker and the advantage of a thicker wire mitigated. Conversely, a thin wire results in a higher electron velocity at the end of their drift, and therefore a higher induced current ( $I(r) \simeq v_e(r) \times E_w(r) $). Since the experimental noise is relatively constant, there is therefore an increase in the signal-to-noise ratio which could explain the improvement in the energy resolution.

The obtained results confirm that a CPC equipped with a non-polarized anode, combined with the ionization mode, offers very attractive optimization perspectives, whether in terms of polarization voltage, signal-to-noise ratio, or even anode thickness. Compared to the use of a detector equipped with a polarized anode where the noise strongly depends on the voltage (or a PPC which requires very strong polarization), the proposed CPC already offers more robust operating conditions.
There is a balance to be found in the choice of detector parameters, and a reduction of the electronic noise could result in a further improvement in the energy resolution.

\section{Conclusions}
A cylindrical time projection chamber, equipped with a single anode
and operating at high pressure, appears to combine many of the
prerequisites necessary for a detector dedicated to
the search for rare phenomena.
In particular, one of the key points was to demonstrate that an energy resolution of the order of a percent could be achieved.
In the present work the energy resolution was measured with $\alpha$'s of about 5~MeV
in argon and xenon at pressures spanning between 1 and
10 bar. 
The presented observations validated a percent level energy
resolution independently on the gas pressure and type, and on
the initial deposited energy radial position.
These achievements are also exciting to the extent that, with such a detector, detection
adjustments could be made first under argon, before using xenon for the physics program.
It is also worth mentioning that all the developments presented could be quite
easily transposed into a CPC filled with liquid argon or liquid xenon.
Finally, the obtained results pave the way to our final step in detector R\&D for the search of the neutrinoless double beta decay program: designing a detector with the lowest possible material budget~\cite{LautridouXesat,FabriceTPC}.

\begin{acknowledgements}
The authors would like to thank the IdEx Bordeaux 2019 Emergence program for the OWEN grant for the ``Development of a custom made electronics for a single channel time projection chamber detector aiming at the discovery of neutrinoless double beta decays, and for possible applications in industry''. We thank the LP2I Bordeaux and SUBATECH technical staff. We are very grateful to Jordan McElwee for the careful reading of the paper and for his corrections.
\end{acknowledgements}

\bibliographystyle{spphys}       
\bibliography{R2D2bibliography}   

\begin{thebibliography}{10}
\providecommand{\url}[1]{{#1}}
\providecommand{\urlprefix}{URL }
\expandafter\ifx\csname urlstyle\endcsname\relax
  \providecommand{\doi}[1]{DOI \discretionary{}{}{}#1}\else
  \providecommand{\doi}{DOI \discretionary{}{}{}\begingroup
  \urlstyle{rm}\Url}\fi

\bibitem{LEGEND:2022bzq}
N.~Burlac and G.~Salamanna, Nucl. Instrum. Meth. A \textbf{1048}, 167943
  (2023).
\newblock \doi{10.1016/j.nima.2022.167943}

\bibitem{CUORE:2022uex}
I.~Nutini et~al., Int. J. Mod. Phys. A \textbf{37}(07), 2240014 (2022).
\newblock \doi{10.1142/S0217751X22400140}

\bibitem{nEXO:2017nam}
J.~B. Albert et~al., Phys. Rev. C \textbf{97}(6), 065503 (2018).
\newblock \doi{10.1103/PhysRevC.97.065503}

\bibitem{Next:2019}
{Gomez-Cadenas J.J., Monrabal Capilla F. and Ferrario P.}, Front. Phys.
  \textbf{7:51} (2019).
\newblock \doi{10.3389/fphy.2019.00051}

\bibitem{Giomataris:2008ap}
I.~Giomataris et~al., JINST \textbf{3}, P09007 (2008).
\newblock \doi{10.1088/1748-0221/3/09/P09007}

\bibitem{Giomataris:2003bp}
Y.~Giomataris and J.~D. Vergados, Nucl. Instrum. Meth. A \textbf{530}, 330
  (2004).
\newblock \doi{10.1016/j.nima.2004.04.223}

\bibitem{Giomataris:2005fx}
Y.~Giomataris and J.~D. Vergados, Phys. Lett. B \textbf{634}, 23 (2006).
\newblock \doi{10.1016/j.physletb.2006.01.040}

\bibitem{Gerbier:2014jwa}
G.~Gerbier et~al., astro-ph.IM/1401.7902  (2014)

\bibitem{NEWS-G:2017pxg}
Q.~Arnaud et~al., Astropart. Phys. \textbf{97}, 54 (2018).
\newblock \doi{10.1016/j.astropartphys.2017.10.009}

\bibitem{Savvidis:2016wei}
I.~Savvidis, I.~Katsioulas, C.~Eleftheriadis, I.~Giomataris, and
  T.~Papaevangellou, Nucl. Instrum. Meth. A \textbf{877}, 220 (2018).
\newblock \doi{10.1016/j.nima.2017.09.014}

\bibitem{Meregaglia:2017nhx}
A.~Meregaglia et~al., JINST \textbf{13}(01), P01009 (2018).
\newblock \doi{10.1088/1748-0221/13/01/P01009}

\bibitem{Bouet:2020lbp}
R.~Bouet et~al., JINST \textbf{16}(03), P03012 (2021).
\newblock \doi{10.1088/1748-0221/16/03/P03012}

\bibitem{Bouet:2022kav}
R.~Bouet et~al., Nucl. Instrum. Meth. A \textbf{1028}, 166382 (2022).
\newblock \doi{10.1016/j.nima.2022.166382}

\bibitem{Bouet:2023zyk}
R.~Bouet et~al., JINST \textbf{18}(10), T10001 (2023).
\newblock \doi{10.1088/1748-0221/18/10/T10001}

\bibitem{COMSOL}
COMSOL, COMSOL Multiphysics {\textregistered} v. 6.1. www.comsol.com COMSOL AB,
  Stockholm, Sweden  (2022)

\bibitem{EDELWEISS:2017lvq}
E.~Armengaud et~al., JINST \textbf{12}(08), P08010 (2017).
\newblock \doi{10.1088/1748-0221/12/08/P08010}

\bibitem{Plante:2022khm}
G.~Plante, E.~Aprile, J.~Howlett, and Y.~Zhang, Eur. Phys. J. C
  \textbf{82}(10), 860 (2022).
\newblock \doi{10.1140/epjc/s10052-022-10832-w}

\bibitem{Akimov:2017gxm}
D.~Y. Akimov et~al., Instrum. Exp. Tech. \textbf{60}(6), 782 (2017).
\newblock \doi{10.1134/S002044121705013X}

\bibitem{BOLOTNIKOV1996619}
A.~Bolotnikov and B.~Ramsey, Nucl. Instrum. Meth. A \textbf{383}(2), 619
  (1996).
\newblock \doi{https://doi.org/10.1016/S0168-9002(96)00752-8}

\bibitem{Dobi:2010ai}
A.~Dobi, D.~S. Leonard, C.~Hall, L.~Kaufman, T.~Langford, S.~Slutsky, and Y.~R.
  Yen, Nucl. Instrum. Meth. A \textbf{620}, 594 (2010).
\newblock \doi{10.1016/j.nima.2010.03.151}

\bibitem{Vignoli:2015jxa}
C.~Vignoli, Phys. Procedia \textbf{67}, 796 (2015).
\newblock \doi{10.1016/j.phpro.2015.06.135}

\bibitem{XENON:2020fbs}
E.~Aprile et~al., Eur. Phys. J. C \textbf{81}(4), 337 (2021).
\newblock \doi{10.1140/epjc/s10052-020-08777-z}

\bibitem{XENON100:2017gsw}
E.~Aprile et~al., Eur. Phys. J. C \textbf{77}(6), 358 (2017).
\newblock \doi{10.1140/epjc/s10052-017-4902-x}

\bibitem{Veenhof:1993hz}
R.~Veenhof, Conf. Proc. C \textbf{9306149}, 66 (1993)

\bibitem{Veenhof:1998tt}
R.~Veenhof, Nucl. Instrum. Meth. A \textbf{419}, 726 (1998).
\newblock \doi{10.1016/S0168-9002(98)00851-1}

\bibitem{Baibussinov:2009gs}
B.~Baibussinov et~al., JINST \textbf{5}, P03005 (2010).
\newblock \doi{10.1088/1748-0221/5/03/P03005}

\bibitem{LePort:2011hy}
F.~LePort et~al., Rev. Sci. Instrum. \textbf{82}, 105114 (2011).
\newblock \doi{10.1063/1.3653391}

\bibitem{Ramo:1939vr}
S.~Ramo, Proc. Ire. \textbf{27}, 584 (1939).
\newblock \doi{10.1109/JRPROC.1939.228757}

\bibitem{Shockley:1938itm}
W.~Shockley, J. Appl. Phys. \textbf{9}(10), 635 (1938).
\newblock \doi{10.1063/1.1710367}

\bibitem{Dris:2014qpa}
M.~Dris and T.~Alexopoulos, hep-ex/1406.3217  (2014)

\bibitem{Recine:2014zca}
K.~A. Recine, J.~B.~R. Battat, and S.~Henderson, Am. J. Phys. \textbf{82}, 322
  (2014).
\newblock \doi{10.1119/1.4864642}

\bibitem{Warburton:2004nql}
W.~K. Warburton, B.~Dwyer-McNally, M.~Momayezi, and J.~E. Wahl, in \emph{{2004
  IEEE Nuclear Science Symposium and Medical Imaging Conference}} (2004), 1,
  pp. 577--581.
\newblock \doi{10.1109/NSSMIC.2004.1462261}

\bibitem{Lautridou:2023vrk}
P.~Lautridou, J. Phys. Conf. Ser. \textbf{2502}(1), 012006 (2023).
\newblock \doi{10.1088/1742-6596/2502/1/012006}

\bibitem{Sauli77}
F.~Sauli, CERN Yellow Reports: Monographs. \textbf{CERN-77}(09) (1977).
\newblock \doi{10.5170}

\bibitem{Schwegler2014}
P.~Schwegler, {High-Rate Performance of Muon Drift Tube Detectors, Doctoral
  dissertation, Fakultät für Physik der Technischen Universität München}
  (2014)

\bibitem{LautridouXesat}
P.~Lautridou, 6th International Workshop on Application of Noble Gas Xenon to
  Science and Technology  (2023)

\bibitem{FabriceTPC}
F.~Piquemal, 11th international symposium on Large TPCs for low-energy rare
  event detection  (2023)

\end{thebibliography}

\end{document}